\documentclass[aps,prd,floats,floatfix,twocolumn,nofootinbib,amssymb]{revtex4}

\usepackage{graphicx}
\usepackage{bm}

\newcommand{\third}{{\scriptstyle\frac{1}{3}}}

\newcommand{\eqref}[1]{(\ref{#1})}
\newcommand{\beq}{\begin{equation}}
\newcommand{\eeq}{\end{equation}}
\newcommand{\beqa}{\begin{eqnarray}}
\newcommand{\eeqa}{\end{eqnarray}}

\newcommand{\three}[1]{{^{(3)}} #1 }

\newcommand{\gammaHone}{\gamma^{{}_H}_1}
\newcommand{\gammaHtwo}{\gamma^{{}_H}_2}

\begin{document}

\title{Gauge Drivers for the Generalized Harmonic Einstein Equations}

\author{Lee Lindblom$^{1}$, Keith D. Matthews$^{1}$, Oliver Rinne$^{1,2,3}$, 
  and Mark A. Scheel$^{1}$}

\affiliation{$^{1}$Theoretical Astrophysics 130-33, California Institute of
Technology, Pasadena, CA 91125}

\affiliation{$^{2}$Department of Applied Mathematics and Theoretical
  Physics, Centre for Mathematical Sciences, Wilberforce Road,
  Cambridge CB3 0WA, UK}

\affiliation{$^{3}$King's College, Cambridge CB2 1ST, UK}

\begin{abstract}
The generalized harmonic representation of Einstein's equation is
manifestly hyperbolic for a large class of gauge conditions.
Unfortunately most of the useful gauges developed over the past
several decades by the numerical relativity community are incompatible
with the hyperbolicity of the equations in this form.  This paper
presents a new method of imposing gauge conditions that preserves
hyperbolicity for a much wider class of conditions, including as
special cases many of the standard ones used in numerical relativity:
e.g., $K$-freezing, $\Gamma$-freezing, Bona-Mass\'o slicing, conformal
$\Gamma$-drivers, etc.  Analytical and numerical results are presented
which test the stability and the effectiveness of this new gauge
driver evolution system.
\end{abstract}

\pacs{04.25.D-, 04.20.Cv, 02.60.Cb, 04.25.dg}

\date{\today}

\maketitle

\section{Introduction}

The gauge (or coordinate) degrees of freedom in the generalized
harmonic (GH) form of the Einstein equations are determined by specifying
the gauge source functions $H^a$.  These functions are defined as the
results of the covariant scalar-wave operator acting on each of the
spacetime coordinates $x^a$:
\begin{equation}
 H^a=\nabla^c\nabla_c\, x^a.
\label{e:GaugeSourceDef}
\end{equation}
The GH form of the Einstein equations can be represented (somewhat
abstractly) as
\begin{equation}
\psi^{cd}\partial_c\partial_d \psi_{ab} + \partial_a H_b + \partial_b H_a=
Q_{ab}(H,\psi,\partial\psi),
\label{e:Einstein}
\end{equation}
where $\psi_{ab}$ is the spacetime metric, $H_a = \psi_{ab}H^b$, 
and $Q_{ab}$ represents lower order terms that depend on $H_a$, the
metric, and its first derivatives.

The GH form of the Einstein equation is manifestly hyperbolic
whenever $H_a$ is specified as an explicit function of the coordinates
and the metric: $H_a = H_a(x,\psi)$.  In this case the terms
$\partial_a H_b$ that appear in Eq.~(\ref{e:Einstein}) contain at most
first derivatives of the metric.  The Einstein equations become,
therefore, a set of second-order wave equations for each component of
the spacetime metric:
\begin{equation}
\psi^{cd}\partial_c\partial_d\psi_{ab}=\hat Q_{ab}(x,\psi,\partial\psi).  
\end{equation}
Thus the Einstein equations are manifestly hyperbolic for any $H_a =
H_a(x,\psi)$.

Most of the useful gauge conditions developed by the numerical
relativity community over the past several decades can not,
unfortunately, be expressed in the simple form $H_a =H_a(x,\psi)$
(unless the full spacetime metric $\psi_{ab}=\psi_{ab}[x]$ is known
{\it a priori}\hspace{0.1em}).  Many of these conditions 
(e.g., maximal slicing or $\Gamma$-drivers)
would require gauge source
functions that depend on the spacetime metric and its first
derivatives: $H_a=H_a(x,\psi,\partial\psi)$.  In this case the terms
$\partial_aH_b$ in Eq.~(\ref{e:Einstein}) depend on the second
derivatives of the metric, $\psi_{ab}$, and this (generically) destroys
the hyperbolicity of the system.

Pretorius~\cite{Pretorius2005c,Pretorius2005a,Pretorius2006} proposed
a way to expand significantly the class of allowed gauge
conditions, by elevating $H_a$ to the status of an independent
dynamical field.  A separate gauge driver equation is introduced to
evolve $H_a$, for example,
\begin{equation}
\nabla^c\nabla_c H_a = Q_a(x,H,\partial H, \psi,\partial\psi),
\label{e:GaugeDriver}
\end{equation}
where $\nabla^c\nabla_cH_a$ denotes the wave operator\footnote{We 
define exactly what we mean by this wave
operator in Sec.~\ref{s:FirstOrderForm}.} acting on $H_a$.  
This gauge driver equation is solved
together with the GH Einstein equations to determine $\psi_{ab}$ and
$H_a$ simultaneously.  In the combined evolution system, consisting of
Eqs.~(\ref{e:Einstein}) and (\ref{e:GaugeDriver}), the $\partial_a
H_b$ terms in Eq.~(\ref{e:Einstein}) are now lower-derivative terms
that do not affect the hyperbolicity of the system.  Thus the combined
GH Einstein plus gauge driver system is manifestly hyperbolic so long
as $Q_a$ on the right in Eq.~(\ref{e:GaugeDriver}) depends only on the
fields and their first derivatives: $Q_a=Q_a(x,H,\partial
H,\psi,\partial\psi)$.  Each of the solutions, $H_a=H_a(x) $, to these
gauge driver equations is a gauge condition. So the gauge driver
system provides a way for $H_a$ to be determined by the metric
and its derivatives in a flexible way without destroying the
hyperbolicity of the GH Einstein equations.

Pretorius used a particular gauge driver equation of
this form to determine $H_t$ in his ground breaking binary black-hole
simulations:
\begin{equation}
\nabla^c\nabla_c H_t = \xi_1(1-N)N^{-p} + \xi_2 
\bigl(\partial_t H_t - N^k\partial_k H_t\bigr)N^{-1},
\end{equation}
where in this case $\nabla^c\nabla_c$ is the covariant scalar-wave
operator, $N$ is the lapse, $N^k$ is the shift, and $\xi_1$, $\xi_2$
and $p$ are constants.  For suitable choices of these parameters,
Pretorius found this system to be quite effective in preventing the
lapse from ``collapsing'' toward zero as the system evolves.
Solutions to this gauge driver do not correspond to any of the
traditional gauge conditions of numerical relativity as far as we
know.

In this paper we introduce a new class of gauge driver equations that
are general enough to provide implementations of (almost) all of the
standard gauge conditions used by the numerical relativity community.
This is done by choosing an appropriate ``source'' term $Q_a$ for the
right side of Eq.~(\ref{e:GaugeDriver}).  The idea is to choose $Q_a$
so that solutions $H_a$ evolve quickly toward a target gauge source
function $F_a$.  This $F_a$ is chosen so that strict equality
$H_a=F_a$ corresponds exactly to the gauge condition of interest to
us.  We limit $F_a$ only by assuming that it depends on the spacetime
metric and its first (but not second) derivatives:
$F_a=F_a(x,\psi,\partial\psi)$.  These new gauge driver equations are
introduced in Sec.~\ref{s:GaugeDriverEquations}, and we show there
that the combined GH Einstein plus gauge driver system is symmetric
hyperbolic for any target gauge source function of this allowed form.
In Sec.~\ref{s:SpecificGauge} we present the target gauge functions
$F_a$ corresponding to (many of) the gauge conditions commonly used by
the numerical relativity community, including maximal slicing,
$K$-freezing, Bona-Mass\'o slicing, conformal $\Gamma$-freezing, and
conformal $\Gamma$-drivers.  In Sec.~\ref{s:Stability} we use
analytical methods to analyze the solutions of the new gauge driver
system.  We show in particular that $H_a$ approaches any (time
independent) $F_a$ exponentially for evolutions of the gauge driver
equations on flat space.  We also demonstrate the effectiveness of the
coupled GH Einstein and gauge driver system for the case of small
perturbations of flat space using Bona-Mass\'o slicing and one of the
conformal $\Gamma$-driver conditions.  In Sec.~\ref{s:NumericalTests}
we show the effectiveness and stability of our implementation of a
particular choice of Bona-Mass\'o slicing and conformal
$\Gamma$-driver condition using numerical solutions of the full
non-linear equations for perturbed single black-hole spacetimes.  We
summarize and discuss these various results in
Sec.~\ref{s:Discussion}.

\section{Gauge Driver Equations}
\label{s:GaugeDriverEquations}

We begin this section by deriving a system of gauge driver equations
in Sec.~\ref{s:Motivation}, and then constructing a general
first-order representation of these equations in
Sec.~\ref{s:FirstOrderForm}.  We derive the characteristic fields for
this system and their associated speeds in
Sec.~\ref{s:CharacteristicFields}, and show that the coupled
gauge-driver and GH Einstein system is symmetric hyperbolic.  We
analyze the constraints in Sec.~\ref{s:Constraints}, and derive
constraint preserving boundary conditions for the gauge-driver fields
in Sec.~\ref{s:BoundaryConditions}.

\subsection{Motivation}
\label{s:Motivation}

In this section we provide some motivation for our choice of gauge
driver equation.  We consider first the case of the gauge driver
$\nabla^c\nabla_c H_a=Q_a$ acting on a fixed flat-space background.
The idea is to choose $Q_a$ so that the solutions, $H_a$, to this
equation quickly approach the desired target gauge source function
$F_a$.  If $F_a$ were constant in space and time, there would be a
fairly obvious and simple choice:
\begin{equation}
\nabla^c\nabla_c H_a=Q_a=\mu^2(H_a-F_a)+2\mu\partial_t H_a,
\label{e:SimpleGaugeDriver}
\end{equation}
where $\mu$ is a freely specifiable constant.  If $H_a$ like $F_a$
were independent of spatial position, 
the gauge driver equation would be equivalent in this
case to the ordinary differential equation,
\begin{equation}
\partial_t^2( H_a-F_a)+2\mu\partial_t( H_a-F_a) + \mu^2(H_a-F_a)=0,
\end{equation}
whose solution has the form: $H_a(t) = F_a +[H_a(0)-F_a]e^{-\mu t}$.
A similar argument applied to the spatial Fourier transform of
Eq.~(\ref{e:SimpleGaugeDriver}) shows that spatially inhomogeneous $H_a$ also
approach $F_a$ exponentially in time.  In this special case (i.e.,
spatially homogeneous and time independent $F_a$) the simple gauge
driver has the desired behavior: all the solutions $H_a$ approach the
target gauge source $F_a$ exponentially on the adjustable time scale
$1/\mu$.

This simple gauge driver, Eq.~(\ref{e:SimpleGaugeDriver}), fails
unfortunately even in flat space if $F_a$ is a generic function of
position. An easy way to see this is to assume that all the solutions
$H_a$ do approach $F_a$ asymptotically as $t\rightarrow\infty$.
Since $F_a$ and consequently $H_a$ are independent of time in this
limit, Eq.~(\ref{e:SimpleGaugeDriver}) reduces to $\nabla^k\nabla_k
H_a=0$, where $\nabla^k\nabla_k H_a$ represents the spatial Laplacian
of $H_a$.  But this is impossible because $\nabla^k\nabla_k F_a\ne 0$
for generic $F_a$.  So (not surprisingly) the simple gauge driver
fails in general.

This gauge driver can be modified in a fairly straightforward way,
however, that corrects this problem.  Define an auxiliary dynamical
field $\theta_a$:
\begin{equation}
\partial_t \theta_a+\eta \theta_a = \nabla^k\nabla_k H_a.
\label{e:SimpleThetaEquation}
\end{equation}
This equation can be integrated analytically to obtain an equivalent
integral representation of $\theta_a$:
\begin{equation}
\theta_a = \theta_a(0)e^{-\eta t}+\int^{t}_0\!\!\! 
e^{-\eta(t-t')}\,\nabla^k\nabla_kH_a(t')\,dt'.
\end{equation}
Thus $\theta_a$ represents an exponentially weighted (in favor of
times near $t$) time average of the past evolution of the term on the
right side of Eq.~(\ref{e:SimpleThetaEquation}).  We can use this
$\theta_a$ to construct an improved gauge driver:
\begin{equation}
\nabla^c\nabla_c H_a=Q_a=\mu^2(H_a-F_a)+2\mu\partial_t H_a+\eta\theta_a.
\label{e:BetterGaugeDriver}
\end{equation}
If a solution to the improved gauge driver approaches a time
independent state, then Eq.~(\ref{e:SimpleThetaEquation}) implies that
$\eta\theta_a=\nabla^k\nabla_k H_a$.
Equation~(\ref{e:BetterGaugeDriver}) reduces in this case to
$0=\mu^2(H_a-F_a)$.  So the addition of the time averaging field
$\theta_a$ forces $H_a$ to approach $F_a$ in any time independent
state, even in the case of inhomogenous $F_a$.  The remainder of this
paper is devoted to the analysis of this improved gauge driver,
Eq.~(\ref{e:BetterGaugeDriver}), suitably generalized for use in an
arbitrary spacetime.

\subsection{First Order Form}
\label{s:FirstOrderForm}

We find it very useful to consider the first-order representations of
evolution systems (such as our gauge driver) for a variety of reasons:
from basic mathematical issues (such as the formulation of appropriate
boundary conditions), to more practical code stability issues.  This
section develops a first-order representation of the gauge driver
system, suitably generalized for use in an arbitrary curved spacetime.

The gauge driver system described above evolves $H_a$ through a wave
equation of the form $\nabla^c\nabla_c H_a=Q_a$.  In a general curved
spacetime, we assume that $\nabla^c\nabla_cH_a$ represents the
covariant wave operator that treats $H_a$ as a co-vector.  This choice
needs a bit of clarification, since the gauge source function $H_a$ is
not actually a co-vector.  One way of giving meaning to this equation
is to use the gauge driver system to determine a new field $\tilde H_a$
that does transform as a co-vector: $\nabla^c\nabla_c \tilde H_a = Q_a$.
Then fix $H_a$ by setting $H_a=\tilde H_a$ in some particular
coordinate frame.  This construction is not covariant, but fixing
coordinate conditions can never be completely covariant.  An
equivalent way to do this is to write out and impose the
gauge driver equation, $\nabla^c\nabla_c H_a = Q_a$, only in the
special coordinate frame in which $H_a=\tilde H_a$.  We adopt this
second approach since it simplifies the notation somewhat.  So
throughout this paper the gauge driver equation, $\nabla^c\nabla_c H_a
= Q_a$, will only be imposed in some particular coordinate frame that
we must specify.  In our code we use a global Cartesian coordinate system,
and we will always impose the gauge driver equation in that frame.

Before we discuss the first-order form of the gauge driver equations,
we also need to examine the somewhat pathological covariant vector
wave operator in more detail.  This operator acting on $H_a$ (assumed here
to be a co-vector as discussed above) can be
written out more explicitly in the form:
\begin{eqnarray}
\nabla^c\nabla_c H_a &=& \psi^{bc}\partial_b\partial_c H_a
-\Gamma^{b}\partial_b H_a -2\psi^{bc}\Gamma^d_{ac}\partial_bH_d\nonumber\\
&&+(R_a{}^b-\partial_a\Gamma^b)H_b,
\end{eqnarray}
where $\Gamma^a{}_{bc}$ is the Christoffel connection,
$\Gamma^a=\psi^{bc}\Gamma^a{}_{bc}$, and $R_a{}^b$ is the associated
Ricci curvature.  This wave operator is well behaved on a fixed
background spacetime.  However the $H_b\partial_a\Gamma^b$ term
includes second derivatives of the metric that would interfere with
hyperbolicity, if it were coupled in a non-trivial way to
the full Einstein equations.  Fortunately this problem has a simple
solution.  Since we use the GH form of the Einstein equations, this
term can be transformed into the more benign form, $-H_b\partial_aH^b$
(or if a more linear looking form is preferred
$\Gamma_b\partial_aH^b$), using the gauge constraint $H^a=-\Gamma^a$ 
~\cite{Lindblom2006}. 
We regard the Ricci tensor $R_a{}^b$ as being determined by the matter
sources via the Einstein equations; in particular, it does not contain
any second derivatives of the metric.
For notational convenience we introduce the quantity $W_a(H)$,
\begin{eqnarray}
W_a(H)&=&2\psi^{bc}\Gamma^d{}_{ac}\partial_bH_d
-(\partial_aH^b+R_a{}^b)H_b,
\end{eqnarray}
that represents the parts of the vector wave operator that are
not present in the scalar wave operator.  Our representation
of the covariant vector wave operator is therefore given by,
\begin{eqnarray}
\nabla^c\nabla_c H_a &=& \psi^{bc}\partial_b\partial_c H_a
-\Gamma^{b}\partial_b H_a -W_a(H).
\end{eqnarray}

To represent this equation in first-order form, we introduce the usual
additional first-order fields $\Pi^H_a$ and $\Phi^H_{ia}$
representing (up to the addition of constraints) 
the appropriate time and space derivatives of $H_a$
respectively:
\begin{eqnarray}
  \Pi^H_a &=& -t^b \partial_b H_a, \\ 
  \Phi^H_{ia} &=& \partial_i H_a .
\end{eqnarray}
Here (and throughout this paper) $t^a$ is the future directed unit normal 
to the $t=$ constant hypersurfaces;
Latin indices $a$ through $h$ are spacetime indices and run from 0 to 3;
and Latin indices 
$i$ through $n$ are spatial indices and run from 1 to 3.
We also define the spatial metric on the $t=$ constant hypersurfaces,
\begin{equation}
  g_{ab} = \psi_{ab} + t_a t_b .
\end{equation}
The covariant wave operator, $\nabla^c\nabla_c H_a$, can then be
expressed in terms of these first-order field variables:
\begin{eqnarray}
\nabla^c\nabla_c H_a &=& t^c\partial_c \Pi^{H}_a + g^{ij}\partial_i\Phi^H_{ja}
-t_b\Gamma^b\Pi^H_a - \Gamma^i\Phi^H_{ia} \nonumber\\
&&+\frac{1}{2} \Pi^H_at^bt^c \Pi_{bc}+ g^{ij}\Phi^H_{ia}t^b \Pi_{bj} \nonumber\\
&&-W_a(H),
\end{eqnarray}
where $W_a(H)$ can be written as 
\begin{eqnarray}
W_a(H)&=&(t_a\Pi_{bc}+g_a{}^i\Phi_{ibc})t^b\psi^{cd}
(\Pi^H_d-t^eH_dH_e)\nonumber\\
&&+(t_a\Pi_{ib}+g_a{}^j\Phi_{jib})\psi^{bc}g^{ik}(\Phi^H_{kc}+H_kH_c)
\nonumber\\
&&-g^{ij}t^b\Pi_{ia}\Phi^H_{jb}
+g^{ij}\psi^{bc}\Phi_{iab}\Phi^H_{jc}\nonumber\\
&&-g^{ij}\Pi^H_j(\Pi_{ai}
+t^b \Phi_{iab})-g^{ij}g^{kl}\Phi_{ika}\Phi^H_{lj}
\nonumber\\
&&+(t_a\Pi^H_b+g_a{}^i\Phi^H_{ib})\Gamma^b-R_a{}^bH_b.
\end{eqnarray}
We note that leaving out the $W_a(H)$ terms is equivalent to applying the
covariant scalar wave operator to each component of $H_a$ in our
special coordinate frame.  We also
note that the remaining $\Gamma^b$ terms that appear in the above
equations are to be thought of as functions of the first-order GH
fields:
\begin{eqnarray}
\label{e:GammaInGHVars1}
\Gamma^b &=& \psi^{bc}t^d\Pi_{cd}+g^{ij}\psi^{bc}\Phi_{ijc}
-\frac{1}{2}\psi^{cd}(t^b\Pi_{cd}+g^{bi}\Phi_{icd}).
\nonumber\\
\end{eqnarray}

The representation of wave equations of this type in
first-order form is well understood, see e.g., Refs.~\cite{Holst2004,
Lindblom2006}; the result for our gauge driver equation is
\begin{eqnarray}
&&\partial_t H_a - (1+\gammaHone)N^k\partial_k H_a 
= -N \Pi^{H}_a
-\gammaHone N^k\Phi^{H}_{ka},\nonumber\\
\label{e:psidot}\\
&&\partial_t \Pi^{H}_a  - N^k\partial_k \Pi^H_a  +Ng^{ki}\partial_k \Phi^H_{ia}
-\gammaHone\gammaHtwo
N^k\partial_kH_a=\nonumber\\
&&\qquad\qquad-\gammaHone
\gammaHtwo N^k\Phi^H_{ka}+
N J^k \Phi^H_{ka} + NK \Pi^H_a\nonumber\\
&&\qquad\qquad + Q_a+N W_a,
\label{e:pidot}\\
&&\partial_t \Phi^H_{ia} - N^k\partial_k \Phi^H_{ia} + N\partial_i \Pi^H_a 
- \gammaHtwo N\partial_i H_a
=\!\nonumber\\
&&\qquad\qquad
- \Pi^H_a\partial_iN+ \Phi^H_{ka}\partial_iN^k
-\gammaHtwo N \Phi^H_{ia}.\label{e:Phidotnew}
\end{eqnarray}
The quantities $N$, $N^k$, and $g_{ij}$ that appear in these equations
are the lapse, shift and spatial metric, defined by the usual
three-plus-one representation of the spacetime metric:
\begin{eqnarray}
ds^2 &=& \psi_{ab}dx^adx^b\nonumber\\
     &=&-N^2 dt^2 + g_{ij}(dx^i + N^idt)(dx^j + N^jdt).\qquad
\label{e:threeplusonemetric}
\end{eqnarray}
The auxiliary quantities $K$ and $J^i$ in Eq.~(\ref{e:pidot}) depend
on the background spacetime geometry and can be written in terms of
the first-order GH Einstein variables:
\begin{eqnarray}
K&=& \frac{1}{2} g^{ij} \Pi_{ij} + g^{ij} t^a \Phi_{ija},\label{e:Kdef}\\ 
J^i&=&
\left(g^{jk} g^{li}-\frac{1}{2} g^{ij}g^{kl}\right) \Phi_{jkl} + \frac{1}{2}
g^{ij} t^a t^b \Phi_{jab} .\label{e:Jdef}\qquad
\end{eqnarray}
The constants $\gammaHone$ and $\gammaHtwo$
are introduced (in analogy with the first-order GH Einstein
system~\cite{Lindblom2006}) to allow us to control the growth of
constraint violations, and to allow us to adjust one of the
characteristic speeds of the system.

The quantity $Q_a$ in Eq.~(\ref{e:pidot}) is defined by a natural 
generalization of Eq.~(\ref{e:BetterGaugeDriver}):
\begin{eqnarray}
Q_a&=&\mu^2_1(1-\xi_1)N (H_a-F_a)\nonumber\\
&&-2\mu_2(1-\xi_2) N \Pi^H_a +\eta_1\theta_a.
\label{e:definitionQ}
\end{eqnarray}
The differences between this expression and
Eq.~(\ref{e:BetterGaugeDriver}) are an overall factor of the lapse $N$
(to convert from coordinate time to proper time), the replacement of
$\partial_t H_a$ by $\Pi^H_a$ (the first order field representing
$-t^c\partial_c H_a$), the introduction of independent damping
parameters $\mu_1$ and $\mu_2$, and the introduction
of new constant parameters $\xi_1$ and $\xi_2$.  
The purpose of these latter parameters, $\xi_1$ and $\xi_2$, is to
move the damping terms (or fractions thereof)
into the source for the time-averaging field $\theta_a$ 
(cf. Eq.~(\ref{e:thetadef}) below), thus effectively
replacing these terms by their time averages.
We assume as before that $F_a$ is a given
function of the four-metric and its first derivatives:
$F_a=F_a(x,\psi,\partial\psi)$.

The evolution equation for $\theta_a$ is chosen, in analogy with
Eq.~(\ref{e:SimpleThetaEquation}), to include as its source all the
terms in Eq.~(\ref{e:pidot}) that do not vanish automatically in a
time independent state:
\begin{eqnarray}
&&\!\!\!\!\partial_t\theta_a +\eta_1\theta_a
=2\mu_2[\xi_3N\Pi^H_a+(1-\xi_3)(1+\gammaHone) N^k\partial_k H_a]
\nonumber\\
&&\qquad \qquad\quad+ Ng^{ki}\partial_k \Phi^H_{ia}
-N^k\partial_k \Pi^H_a -\gammaHone
\gammaHtwo N^k\partial_k H_a\nonumber\\
&&\qquad\qquad\quad
-2\mu_2(1-\xi_3)\gammaHone N^k \Phi^H_{ka}
+\gammaHone\gammaHtwo 
N^k \Phi^H_{ka}\nonumber\\
&&\qquad\qquad\quad
-N K \Pi^H_a- N J^i \Phi^H_{ia}-NW_a\nonumber\\
&&\qquad\qquad\quad
+\mu^2_1\xi_1 N(H_a-F_a)-2\mu_2\xi_2N\Pi^H_a.
\label{e:thetadef}
\end{eqnarray}
The $\xi_3$ parameter is introduced to add a multiple of $\partial_t H_a$
to the source of the time averaging field.  We use Eq.~(\ref{e:psidot}) to
re-express this $\partial_t H_a$ as the terms proportional to $\xi_3$ that
appear on the right side of Eq.~(\ref{e:thetadef}).
Assuming the system approaches a state in
which $H_a$ becomes time independent, then $\theta_a$ exponentially
approaches the time independent limit of the terms on the right side
of Eq.~(\ref{e:thetadef}).  These terms were chosen so that
Eq.~(\ref{e:pidot}) then implies that $H_a\rightarrow F_a$ in this
limit.
Our choices for the parameters, $\mu_1$, $\mu_2$, $\eta_1$, $\xi_1$,
$\xi_2$ and $\xi_3$ that appear in
Eqs.~(\ref{e:definitionQ}) and~(\ref{e:thetadef}) will be guided
by the stability analysis that we perform in
Sec.~\ref{s:Stability}.  

\subsection{Characteristic Fields}
\label{s:CharacteristicFields}

The gauge driver evolution Eqs.~(\ref{e:psidot})--(\ref{e:Phidotnew})
and (\ref{e:thetadef}) comprise a first-order evolution system of the
form,
\begin{equation}
\partial_t u^\alpha + A^{k\,\alpha}{}_\beta \partial_k u^\beta = B^\alpha,
\label{e:linearsystem}
\end{equation}
for the fields $u^\alpha= \{H_a,\Pi^H_a,\Phi^H_{ia},\theta_a\}$
(treating the spacetime metric for the moment as a fixed background
field).  The characteristic fields of such an evolution system are
important for a number of reasons, including the formulation of outer
boundary conditions and exchanging information across internal
boundaries of the computational domain.  The characteristic fields (in
the direction of a unit spacelike covector $n_k$) are defined as the
projections of the fields $u^\alpha$ onto the left eigenvectors of the
characteristic matrix $n_kA^{k\,\alpha}{}_\beta$.  For the gauge
driver system, these characteristic fields are
\begin{eqnarray}
U^{H\pm}_a&=&\Pi^H_a\pm n^i \Phi^H_{ia}-\gammaHtwo H_a,\\
Z^{H1}_a&=&H_a,\\
Z^{H2}_{ia} &=& P_i{}^j \Phi^H_{ja},\\
Z^{H3}_a&=&\theta_a + \Pi^H_a - 2\mu_2(1-\xi_3) H_a,
\end{eqnarray}
where $P_{ij} \equiv g_{ij} - n_i n_j$. 

The eigenvalues associated with the characteristic fields are called
the characteristic speeds of the system.  For the gauge driver system,
the characteristic fields $U^{H\pm}_a$ have speeds $\pm N - n_i N^i$,
$Z^{H1}_a$ has speed $-(1+\gammaHone)n_i N^i$, $Z^{H2}_{ia}$ has speed
$-n_iN^i$, and $Z^{H3}_a$ has speed zero.

The inverse transformation between dynamical fields and characteristic
fields for our gauge driver system is
\begin{eqnarray}
H_a&=&Z^{H1}_a,\\
\Pi^H_a &=& \frac{1}{2}(U^{H+}_a+U^{H-}_a)+\gammaHtwo Z^{H1}_a,\\
\Phi^H_{ia}&=& \frac{1}{2}(U^{H+}_a-U^{H-}_a)n_i + Z^{H2}_{ia},\\
\theta_a&=& Z^{H3}_a -\frac{1}{2}(U^{H+}_a+U^{H-}_a)\nonumber\\
&&+2\mu_2(1-\xi_3) Z^{H1}_a-\gammaHtwo 
Z^{H1}_a.\quad
\end{eqnarray}
The existence of this inverse transformation shows that there is a one-to-one
correspondence between the dynamical fields and the characteristic fields.
This implies that the gauge driver system is strongly hyperbolic.

A quasi-linear evolution system, Eq.~(\ref{e:linearsystem}), is
symmetric hyperbolic (a stronger condition than strong hyperbolicity)
if there exists a positive definite metric $S_{\alpha\beta}$ 
(called a {\em symmetrizer}) on the
space of fields, such that $S_{\alpha\gamma}A^{k\,\gamma}{}_\beta
=S_{\beta\gamma}A^{k\,\gamma}{}_{\alpha}$.  The gauge driver system,
Eqs.~(\ref{e:psidot})--(\ref{e:Phidotnew}) and (\ref{e:thetadef}),
does have such a symmetrizer:
\begin{eqnarray}
dS^2&=& S_{\alpha\beta}du^\alpha du^\beta\nonumber\\
&=& \sum_a\Bigl\{\Lambda_a^2dH_a^2+
\Bigl[d\theta_a-2\mu_2(1-\xi_3) dH_a+d\Pi^H_a\Bigr]^2 \nonumber\\ 
&&\qquad+g^{ij}d\Phi^H_{ia}d\Phi^H_{ja}+ (d\Pi^H_a-\gammaHtwo dH_a)^2\Bigr\},
\quad
\end{eqnarray}
where $\Lambda_a$ are arbitrary (non-vanishing) constants.  The gauge
driver system is therefore symmetric hyperbolic.

Up to this point the discussion has focused on the properties of the
gauge driver system, Eqs.~(\ref{e:psidot})--(\ref{e:Phidotnew}) and
(\ref{e:thetadef}), with the spacetime metric considered as a fixed
background field.  Our real interest of course is the case where
the gauge driver system is coupled to the GH Einstein system,
Eq.~(\ref{e:Einstein}).  Thus we need to consider the properties of
the combined evolution system having as dynamical fields the gauge
driver fields plus the GH Einstein system fields: $u^\alpha=
\{H_a,\Pi^H_a, \Phi^H_{ia},\theta_a,\psi_{ab},\Pi_{ab},\Phi_{iab}\}$.
The fields $\Pi_{ab}$ and $\Phi_{iab}$ represent the first derivatives
of the spacetime metric $\psi_{ab}$, as defined for example in
Ref.~\cite{Lindblom2006}.  We need to analyze the properties of the
characteristic matrix $A^{k\,\alpha}{}_\beta$ of this combined system
to determine whether the full coupled system is hyperbolic.

We have shown above that the gauge driver system,
Eqs.~(\ref{e:psidot})--(\ref{e:Phidotnew}) and (\ref{e:thetadef}), is
symmetric hyperbolic if the spacetime metric is considered as a
background field.  Similarly, the first-order representation of the GH
Einstein system~\cite{Lindblom2006} is symmetric hyperbolic if the
gauge source function $H_a$ is considered as a background field.  
The characteristic matrix $A^{k\,\alpha}{}_\beta$ of the combined system
is block diagonal, except for any cross terms that might arise if
derivatives of the gauge driver fields appear in evolution equations for the
GH fields or vice versa.  The only potential cross terms
are as follows: The term $2\partial_{(a}H_{b)}$
occurs in the GH Einstein equations~(\ref{e:Einstein}),
the quantities $K$ and $J_i$ in Eq.~(\ref{e:pidot}) depend on
derivatives of the spacetime metric, and the target gauge source function
$F_a$ appearing in Eq.~(\ref{e:pidot}) may include
derivatives of the spacetime metric.

However, $\partial_{(a}H_{b)}$ can be rewritten in terms of the
first-order gauge driver variables as
\begin{equation}
  \partial_{(a}H_{b)}= \Phi{}_{i(a}^Hg{}_{b)}{}^i + \Pi{}_{(a}^H t{}_{b)}.
\end{equation}
%
Likewise, $K$ and $J_i$ can be expressed as algebraic functions
of the first-order GH fields $\psi_{ab}$, $\Pi_{ab}$, and $\Phi_{iab}$,
cf. Eqs.~(\ref{e:Kdef}) and (\ref{e:Jdef}).
Finally, the target gauge source function $F_a$ is assumed to be a function of
the metric and its first derivatives, so it also can
be written as an algebraic function of the first-order fields: $F_a=
F_a(x,\psi,\Pi,\Phi)$.  Thus all of these potential cross terms can be
written as algebraic functions of the dynamical fields and do not
contribute to the characteristic matrix $A^{k \alpha}{}_\beta$ at all.

The characteristic matrix of the combined evolution system is
therefore block diagonal.  It follows that the characteristic fields
of the combined system are just the collection of unmodified
characteristic fields from the separate systems.  Similarly the matrix
$S_{\alpha\beta}$ needed to symmetrize the full system is just the
matrix whose diagonal blocks are the symmetrizers of the individual
systems.  It follows trivially that the combined GH Einstein and gauge
driver system is both strongly and symmetric hyperbolic.

\subsection{Constraints}
\label{s:Constraints}

The basic gauge driver evolution system, Eq.~(\ref{e:GaugeDriver}),
has no fundamental constraints.  However by transforming the system to
first-order form, Eqs.~(\ref{e:psidot})--(\ref{e:Phidotnew}), we
introduce a set of new constraints:
\begin{eqnarray}
{\cal C}{}_{ia}^H&=& \partial{}_i H{}_a - \Phi{}_{ia}^H,\\
{\cal C}{}_{ija}^H &=& 2\partial{}_{[i}{\cal C}{}_{j]a}^H
=-2\partial{}_{[i}\Phi{}_{j]a}^H.
\end{eqnarray}
These constraints vanish, ${\cal C}{}_{ia}^H={\cal
C}{}_{ija}^H=0$, if and only if a solution to the first-order system also
represents a solution to the original second-order equation.

These constraints are determined by the values of the dynamical fields
$H_a$ and $\Phi^H_{ia}$, therefore their time evolution is determined
by the gauge driver evolution system.  It is straightforward to show
that these constraints satisfy the evolution equations
\begin{eqnarray}
\partial_t {\cal C}^H_{ia} &-& (1+\gammaHone)N^k\partial_k{\cal C}^H_{ia} 
= - (1+\gammaHone)\partial_iN^k{\cal C}^H_{ka}\nonumber\\
&&\qquad\qquad\qquad\quad
-\gammaHtwo N {\cal C}^H_{ia} -\gammaHone N^k {\cal C}^H_{kia},
\label{e:HConstEvol1}\\
\partial_t {\cal C}^H_{ija} &-& N^k\partial_k{\cal C}^H_{ija}=
\partial_iN^k{\cal C}^H_{kja}+ \partial_jN^k{\cal C}^H_{ika}
-\gammaHtwo N {\cal C}^H_{ija} \nonumber\\ 
&&\qquad\qquad\qquad\quad- 2\gammaHtwo 
\partial_{[i}N {\cal C}^H_{j]a},
\label{e:HConstEvol2}
\end{eqnarray}
as a consequence of Eqs.~(\ref{e:psidot})--(\ref{e:Phidotnew}).  

The characteristic matrix of this constraint evolution system is
diagonal, so the constraints are themselves characteristic fields of
this system. The constraint ${\cal C}^H_{ia}$ propagates at the speed
$-(1+\gammaHone)n_kN^k$, while ${\cal C}^H_{ija}$ propagates at the
speed $-n_kN^k$.  This constraint evolution system is strongly (and
also symmetric) hyperbolic.  

The constraint evolution system is also homogeneous in the
constraints, i.e., the right sides of Eqs.~(\ref{e:HConstEvol1}) and
(\ref{e:HConstEvol2}) are proportional to the constraints.  This
implies, for example, that these constraints will remain satisfied
within the domain of dependence of the subset of the initial surface
on which they are satisfied.

\subsection{Boundary Conditions}
\label{s:BoundaryConditions}

Boundary conditions are needed for any of the characteristic fields
having incoming (i.e., negative) characteristic speeds on the
boundary.  Some of these boundary conditions can be determined by the
need to prevent the influx of constraint violations, while others
can be chosen to control the particular gauge condition being imposed
at the boundary.
In analogy with the scalar field system~\cite{Holst2004}, the needed
constraint preserving boundary conditions for this system are:
\begin{eqnarray}
d_t Z^{H1}_a &=& D_t Z^{H1}_a - (1 + \gammaHone) n_k N^k n^i {\cal C}^H_{ia},\\
d_t Z^{H2}_{ia} &=& D_t Z^{H2}_{ia} - n_k N^k P_i{}^j n^l{\cal C}^H_{jla},
\end{eqnarray}
where $d_t Z^{H1}_a=\partial_t H_a$ and $d_t Z^{H2}_{ia}
=P_i{}^k\partial_t \Phi^H_{ka}$ represent the constraint field
projections of the time derivatives of the dynamical fields, while
$D_t Z^{H1}_a$ and $D_tZ^{H2}_{ia}$ represent the constraint field
projections of the right sides of the evolution equations for these
fields.

The characteristic fields $U^{H\pm}_a$ need boundary conditions
whenever the corresponding speeds $v_\pm=\pm N - n_kN^k$ are negative.
Since $v_-<v_+$, typically the $U^{H-}_a$ mode is the one needing a
boundary condition.  The boundary condition on this field controls the
incoming part of the gauge condition being imposed on the boundary.
We often use a ``freezing'' boundary condition, $\partial_t
U^{H-}_a=0$, or the boundary condition, $\partial_t
U^{H-}_a=-\gammaHtwo \partial_t Z^{H1}_a$.  Ideally the boundary
condition on the characteristic field $U^{H-}_a$ should be determined
by the gauge condition that the driver equation is trying to enforce,
however at present we do not know how to do this.

\section{Specific Gauge Conditions}
\label{s:SpecificGauge}

The gauge driver equations presented in
Sec.~\ref{s:GaugeDriverEquations} were designed to evolve the
gauge source function, $H_a$, toward a target function
$F_a=F_a(x,\psi,\partial\psi)$.  The question of how well these
equations accomplish this will be explored in Secs.~\ref{s:Stability}
and \ref{s:NumericalTests}. Here we focus on the issue of
constructing target functions $F_a$ for particular gauge conditions
used in numerical relativity.

Most of the gauge choices used by the numerical relativity community,
including all the examples below, are expressed as conditions on the 
spacetime metric and its first (space and time) derivatives;
so abstractly, all such gauge conditions can be written
in the form $G_a(x,\psi,\partial\psi)=0$.  
Whenever the GH Einstein constraints are satisfied, it follows from
Eq.~(\ref{e:GaugeSourceDef}) that
$H_a=-\Gamma_a\equiv-\Gamma_{abc}\psi^{bc}$, where $\Gamma_{abc}$ is
the four-dimensional Christoffel symbol.  An appropriate target gauge
source function $F_a$ is therefore given by
\begin{equation}
F_a=-\Gamma_{a}-\rho\, G_a,
\label{e:TargetSourceFunction}
\end{equation}
where $\rho$ is an arbitrary (non-vanishing) constant.  When the
constraints are satisfied, this equation implies that
$H_a-F_a=\rho\,G_a$.  So if the gauge driver system succeeds in
driving $H_a-F_a\rightarrow 0$, it follows that $G_a\rightarrow 0$ as
well for any $\rho\neq0$.  This $F_a$ has the general form assumed in
the discussions of Sec.~\ref{s:GaugeDriverEquations},
$F_a=F_a(x,\psi,\partial\psi)$, whenever $G_a$ has the form
$G_a=G_a(x,\psi,\partial\psi)$.  Therefore the gauge driver system
with this target $F_a$ should enforce the desired gauge condition
$G_a=0$ asymptotically as the system evolves.

The numerical relativity community traditionally separates gauge
conditions into those that determine the lapse $N$ (often called
slicing conditions) and those that determine the spatial coordinates
through the shift $N^k$.  Expressing $\Gamma_a$ in terms of the
three-plus-one representation of the spacetime metric,
Eq.~(\ref{e:threeplusonemetric}), reveals that different components of
$\Gamma_a$ are naturally related to conditions on the lapse and shift
respectively:
\begin{eqnarray}
\Gamma_{\hat t}&\equiv& t^a\Gamma_a= 
N^{-2}(\partial_t N - N^i\partial_i N)+ K ,\\
\Gamma_i &=& -N^{-2}g_{ij}(\partial_t N^j - N^k\partial_k N^j)
-N^{-1}\partial_i N\nonumber \\
&&+{}^{(3)}\Gamma_{ijk}\,g^{jk} ,
\end{eqnarray}
where ${}^{(3)}\Gamma_{ijk}$ is the Christoffel symbol associated with
the three-metric $g_{ij}$.  We see that $\Gamma_{\hat t}$ depends on
the time derivative of the lapse, and that $\Gamma_i$ depends on the
time derivative of the shift.  It is natural then to impose slicing
conditions using the $F_{\hat t}$ component of the target gauge source
function, and to impose shift conditions through the spatial
components $F_i$.  Once $F_{\hat t}$ and $F_i$ are specified, the
time component $F_t$ is obtained from the identity $F_t=N F_{\hat t}+N^kF_k$.
Finally, we will want to express the target gauge source function in
terms of the first-order GH Einstein system variables
$\{\psi_{ab},\Pi_{ab},\Phi_{iab}\}$, therefore the 
expression for $\Gamma_a$ from Eq.~(\ref{e:GammaInGHVars1}) will be useful:
\begin{equation}
  \label{e:GammaInGHVars}
  \Gamma_a = g^{ij} \Phi_{ija} + t^b \Pi_{ba} - \frac{1}{2} g_a{}^i
  \psi^{bc} \Phi_{ibc} - \frac{1}{2} t_a \psi^{bc} \Pi_{bc}.
\end{equation}

The remainder of this section presents a list of target gauge source
functions, $F_a$, that describe commonly used gauge conditions
in numerical relativity.  Slicing conditions are described
in Sec.~\ref{s:SlicingConditions} and shift conditions are
given in Sec.~\ref{s:ShiftConditions}.

\subsection{Slicing Conditions}
\label{s:SlicingConditions}

One of oldest gauge conditions used in numerical
relativity is maximal slicing~\cite{Smarr78b}, where the trace of the
extrinsic curvature of the $t=$ constant hypersurfaces vanishes: $K=0$.
More generally constant curvature slicings are sometimes used, $K =
K_0$, where $K_0$ is constant on each slice (but may be a specified
function of time).  This gauge condition can be written in the form
$G_{\hat t}=0$, where
\begin{equation}
G_{\hat t} = K_0-K = K_0- \frac{1}{2} g^{ij} \Pi_{ij} - g^{ij} t^c \Phi_{ijc}.
\label{e:ConstCurvatureSlicing}
\end{equation}
Using Eq.~\eqref{e:TargetSourceFunction} with
Eq.~\eqref{e:GammaInGHVars} and \eqref{e:ConstCurvatureSlicing}, we obtain
\begin{eqnarray}
  F_{\hat t} &=& - \frac{1}{2} t^a t^b \Pi_{ab} - \rho_1 K_0 
+\frac{1}{2}(\rho_1-1)g^{ij}\Pi_{ij}\nonumber\\
&&+(\rho_1-1)t^ag^{ij}\Phi_{ija}.
\end{eqnarray}
The choice of the arbitrary slicing gauge parameter, $\rho_1=1$, gives
a very simple expression for the constant curvature target gauge
source function $F_{\hat t}$, but other choices may be more stable
or more effective.

Perhaps the most widely used slicing conditions are various members of
the family introduced by Bona and Mass\'o~\cite{Bona94b}.  These gauge
conditions are evolution equations for the lapse $N$ having the
general form\footnote{The original gauge condition in \cite{Bona94b} 
  contains a derivative along the timelike normal instead of a partial
  time derivative.}
\begin{equation}
  \label{e:BonaMassoPartial}
  \partial_t N= -N^2 f(N) (K - K_0),
\end{equation}
where $f(N)$ is an arbitrary function of the lapse.  The particular
case $f(N) = 2/N$ corresponds to the widely used one-plus-log slicing
conditions~\cite{Balakrishna1996,Alcubierre2002,Campanelli2006a,Baker2006a}.
An expression for the general form of this gauge condition in terms of
the first-order GH fields is given by
\begin{eqnarray}
G_{\hat t}
&=&K_0 -g^{ij}t^a\Phi_{ija}-\frac{1}{2}g^{ij}\Pi_{ij}\nonumber\\
&&-\frac{1}{2f(N)}t^at^b\Pi_{ab}
+\frac{1}{2Nf(N)}N^it^at^b\Phi_{iab}.
  \label{e:OnePlusLog1}
\end{eqnarray}
Using this condition in Eq.~(\ref{e:TargetSourceFunction}) results in
the needed target gauge source functions for these Bona-Mass\'o slicing
conditions:
\begin{eqnarray}
  &&F_{\hat t} 
  = \frac{\rho_1-f(N)}{2f(N)} t^a t^b \Pi_{ab} 
- \frac{\rho_1}{2Nf(N)} N^k t^a t^b \Phi_{kab} \nonumber\\
&&\qquad-\rho_1 K_0+
(\rho_1-1)t^ag^{ij}\Phi_{ija}+\frac{1}{2}(\rho_1-1)g^{ij}\Pi_{ij}.\qquad
  \label{e:OnePlusLogF1}
\end{eqnarray}
The slicing gauge parameter choice $\rho_1=1$ makes this expression
for the Bona-Mass\'o gauge condition particularly simple; however, any
choice with $\rho_1\neq0$ is allowed.

\subsection{Shift Conditions}
\label{s:ShiftConditions}

The simplest shift condition (from our perspective) is referred to as
$\Gamma$-freezing~\cite{Baumgarte99}.  This condition fixes the trace
of the Christoffel symbol associated with the conformal spatial metric
$\tilde g_{ij} = g^\lambda g_{ij}$, where $g \equiv \det g_{ij}$ and
$\lambda$ is a constant. (Often $\lambda$ is chosen to be
$\lambda = -\third$ so that $\det \tilde g_{ij}=1$, but any value is
allowed.) The relevant trace of this conformal connection is defined by
\begin{eqnarray}
\three\tilde\Gamma^i &\equiv&
\three\tilde\Gamma^i{}_{jk}\tilde g^{jk}
= g^{-\lambda}
\left(g^{ik}g^{jl}-\frac{1+\lambda}{2}g^{ij}g^{kl}\right)\Phi_{jkl}.
\nonumber\\
\label{e:GammaTilde}
\end{eqnarray}
The $\Gamma$-freezing shift condition simply requires that
\begin{equation}
  \partial_t \three \tilde \Gamma^i = 0 .
\label{e:BasicGammaFreeze}
\end{equation}
For our purposes this must be translated into a condition on the
spacetime metric and its first derivatives.  This is accomplished by
integrating Eq.~(\ref{e:BasicGammaFreeze}) to obtain $\three \tilde
\Gamma^i =\three \tilde \Gamma^i (0)$, where $\three\tilde\Gamma^i(0)$
is the trace evaluated at the initial time.  This condition can be
expressed in terms of first-order GH fields as
\begin{eqnarray}
G_i = g^\lambda g_{ij} \three\tilde\Gamma^j(0)-
\left(\delta_i{}^lg^{jk}-\frac{1+\lambda}{2}\delta_i{}^jg^{kl}\right)\Phi_{jkl}.
\label{e:GammaFreezingG}
\end{eqnarray}
Using Eq.~(\ref{e:TargetSourceFunction}) and (\ref{e:GammaInGHVars}),
this gauge condition is easily transformed into the needed target gauge
source function:
\begin{eqnarray}
  \label{e:GammaFreezingF}
  F_i &=& 
    \frac{1}{2}\bigl[1-\rho_2(1+\lambda)\bigr] g^{jk} \Phi_{ijk}
    - \frac{1}{2} t^a t^b \Phi_{iab}  - t^a \Pi_{ai}  \nonumber\\
&& -\rho_2 g^\lambda g_{ij} \three \tilde \Gamma^j (0) 
   +(\rho_2-1)g^{jk}\Phi_{jki},
\end{eqnarray}
where the shift gauge parameter $\rho_2\neq0$ can be chosen freely.
As a modest generalization we might also want to consider
$\Gamma$-fixing conditions for which $\three\tilde\Gamma^i$ is
specified as a function of time.  For example we might want to set
$\three \tilde \Gamma^i(t) = \three\tilde \Gamma^i(0)e^{-\mu t}$.
This can be done by replacing $\three \tilde \Gamma^i(0)$ with the
desired $\three \tilde \Gamma^i(t)$ in Eqs.~(\ref{e:GammaFreezingG})
and (\ref{e:GammaFreezingF}).

The most commonly used shift conditions in the numerical relativity
community are the $\Gamma$-driver conditions.  The
simplest of these can be written as the following evolution equations
for the shift~\cite{Campanelli2006a},

\begin{eqnarray}
\partial_t N^i &=& B^i,\label{e:UTBgamma1}\\
\partial_t B^i + \eta_2 B^i &=& 
\nu\partial_t{}^{(3)}\tilde\Gamma^i,
\label{e:UTBgamma2}
\end{eqnarray}
where ${}^{(3)}\tilde\Gamma^i$ is the trace of the conformal spatial
connection, Eq.~(\ref{e:GammaTilde}), and $\nu$ and $\eta_2$ are
adjustable constants.  The parameter $\nu$ is usually set to
$\nu=\frac{3}{4}$ on the basis of causality
arguments~\cite{Alcubierre2002,Campanelli2006a}.  But these arguments
do not apply when the lapse and shift are evolved with the GH Einstein
equations, so we leave $\nu$ as an adjustable parameter. Unfortunately
this shift condition is not of the form
$G_i=G_i(x,\psi,\partial\psi)$, which is required by our gauge driver system,
because the right side of Eq.~(\ref{e:UTBgamma2}) depends on second
derivatives of the spacetime metric.  This particular $\Gamma$-driver
condition, Eqs.~(\ref{e:UTBgamma1}) and (\ref{e:UTBgamma2}), can be
transformed however into the more useful form
\begin{eqnarray}
\partial_t N^i &=& \nu
\bigl[{}^{(3)}\tilde\Gamma^i - \eta_2 \Upsilon^i\bigr],
\label{e:NEWgamma1}\\
\partial_t \Upsilon^i &+& \eta_2 \Upsilon^i 
= {}^{(3)}\tilde\Gamma^i.\label{e:NEWgamma2}
\end{eqnarray}
We note that Eqs.~(\ref{e:NEWgamma1}) and (\ref{e:NEWgamma2}) are
equivalent to Eqs.~(\ref{e:UTBgamma1}) and (\ref{e:UTBgamma2})
when $\eta_2\neq0$.  This can be seen by
differentiating $B^i = \nu[\three\tilde\Gamma^i-\eta_2\Upsilon^i]$
with respect to time to determine that Eq.~(\ref{e:UTBgamma2}) is
equivalent to Eq.~(\ref{e:NEWgamma2}). 

This transformed $\Gamma$-driver condition does not depend on the
second derivatives of the spacetime metric, so it is of the form required
for our gauge driver system.  This $\Gamma$-driver condition,
Eq.~(\ref{e:NEWgamma1}), can be written in terms of the first-order GH
fields as
\begin{eqnarray}
G_i&=&-t^a \Pi_{ai} +\frac{1}{N}\, t^a N^j\Phi_{jai}+
\frac{\nu\eta_2}{N^2} \,g_{ij}\Upsilon^j\nonumber\\ 
&&-\frac{\nu}{N^2g^{\lambda}} 
\left(g_i{}^lg^{jk}-\frac{1+\lambda}{2}g_i{}^jg^{kl}
\right)\Phi_{jkl},\qquad
\label{e:NewGammaDriverGHForm}
\end{eqnarray}
where the auxiliary field $\Upsilon^i$ must be evolved using
Eq.~(\ref{e:NEWgamma2}), and is treated as an independent dynamical
field along with
the GH and gauge driver fields.  The addition of Eq.~(\ref{e:NEWgamma2})
to the evolution system does not affect hyperbolicity. (The combined
system has the additional characteristic fields $\Upsilon^i$, all of
which have characteristic speed zero.)  When evolving
Eqs.~(\ref{e:UTBgamma1}) and~(\ref{e:UTBgamma2}), it is common
practice to set $\partial_t N^i=0$ initially~\cite{Campanelli2006a};
the equivalent condition in our notation is initially
choosing $\eta_2\Upsilon^i={}^{(3)}\tilde\Gamma^i$.  The
target gauge source function $F_i$ for this $\Gamma$-driver is
obtained from Eq.~(\ref{e:NewGammaDriverGHForm}) using
Eqs.~(\ref{e:TargetSourceFunction}) and (\ref{e:GammaInGHVars}):
\begin{eqnarray}
  F_i &=& \left(\frac{\nu\rho_2}{N^2g^{\lambda}}-1\right)
\left(g^{jk} g_i{}^l - \frac{1}{2} g_i{}^j g^{kl}\right) 
\Phi_{jkl}  \nonumber\\ 
&& - \frac{1}{2} t^a t^b \Phi_{iab}- \frac{\rho_2}{N} t^a N^j\Phi_{jai}-
\frac{\nu\eta_2\rho_2}{N^2}\, g_{ij}\Upsilon^j\nonumber\\ 
&&-\frac{\nu\lambda\rho_2}{2N^2g^{\lambda}}g^{jk}\Phi_{ijk}
+(\rho_2-1)t^a\Pi_{ai}.
\label{e:NewGammaDriver}
\end{eqnarray}
The shift gauge parameter $\rho_2$ can be chosen to have any non-zero
value.

\section{Flat Space Stability Analysis}
\label{s:Stability}
The analysis of the gauge driver equations in the previous sections is
concerned with rather general questions, such as: Are the equations
hyperbolic?  What are the appropriate boundary conditions?  How are
particular gauge conditions implemented?  In this section (and the
next) we focus on questions about the stability and effectiveness of
the gauge driver equations, such as: Are the gauge driver evolution
equations stable?  How well do the equations actually drive $H_a$
toward the target gauge source function $F_a$?  In this section we use
(mostly) analytical methods to explore these questions for simple cases
that can be described as linear perturbations of flat spacetime.  We
consider three successively more complicated versions of this flat
spacetime problem: First, we analyze the solutions to the gauge driver
equation with a fixed $F_a$ on a flat background spacetime.  Second,
we generalize this problem by allowing $F_a$ to have a prescribed time
dependence. Third, we analyze the more realistic case of the coupled
gauge driver and GH Einstein systems for linear perturbations of flat
spacetime.   We present this analysis in some detail for the
case of a target $F_a$ representing Bona-Mass\'o slicing
and a $\Gamma$-driver shift condition.

Before we specialize to these three specific problems however, we
first establish some common notation and present the basic equations.
Since we are perturbing about flat spacetime, it is convenient to
decompose the solutions into spatial Fourier basis functions.  Thus we
assume that the spatial dependence of each of the perturbed fields is
$e^{ik_jx^j}$, where $k_j$ is a (constant) wave vector, and $x^j$ are
the spatial Cartesian coordinates.  We assume that the gauge source
function $H_a$, the target function $F_a$ and the time averaging field
$\theta_a$ have the forms $H_a(t,x)=\delta H_a(t) e^{ik_jx^j}$,
$F_a(t,x)=\delta F_a(t) e^{ik_jx^j}$, and $\theta_a(t,x)=\delta
\theta_a(t) e^{ik_jx^j}$.  We also assume that the spacetime metric
$\psi_{ab}$ has the form $\psi_{ab}(t,x)=\eta_{ab}+\delta
\psi_{ab}(t)e^{ik_jx^j}$, where $\eta_{ab}$ is the fixed background
Minkowski metric with $N=1$, $g_{ij}=\delta_{ij}$ and $N^i$ constant.
With these assumptions the linearized gauge driver system,
Eqs.~(\ref{e:psidot})--(\ref{e:Phidotnew}) and (\ref{e:thetadef}), can
be written in the form:
\begin{eqnarray}
&&\partial_t \delta H_a -i\beta k\delta H_a= 
\nonumber\\
&&\qquad\qquad- \delta \Pi^H_a
-\gammaHone N^j(\delta\Phi^H_{ja}-ik_j\delta H_a)
,\label{e:HFlatCase}\\
&&\partial_t \delta \Pi^H_a 
+\bigl[2\mu_2(1-\xi_2)-i\beta k \bigr]\delta \Pi^H_a=
- i k^j \delta \Phi^H_{ja}\nonumber\\
&&\qquad\qquad
+ \eta_1\delta\theta_a
+\mu_1^2(1-\xi_1)(\delta H_a-\delta F_a)\nonumber\\
&&\qquad\qquad
-\gammaHone\gammaHtwo N^j(\delta\Phi^H_{ja}-ik_j\delta H_a),
\label{e:HtFlatCase}\\
&&\partial_t (\delta \Phi^H_{ja} - i k_j \delta H_a)
= \nonumber\\
&&\qquad\qquad
+ik_j\gammaHone N^l(\delta\Phi^H_{la}-ik_l\delta H_a)
\nonumber\\
&&\qquad\qquad
-(\gammaHtwo - i\beta k) (\delta \Phi^H_{ja} - i k_j \delta H_a)
,\qquad\label{e:HxFlatCase}\\
&&\partial_t \delta \theta_a +\eta_1\delta \theta_a= 
\bigl[2\mu_2ik\beta(1-\xi_3)+\mu_1^2\xi_1\bigr]\delta H_a\nonumber\\
&& \qquad\qquad
+\bigl[2\mu_2(\xi_3-\xi_2)-i\beta k\bigr] \delta \Pi^H_a+
i k^j\delta \Phi^H_{ja} \nonumber\\
&&\qquad\qquad
-2\mu_2(1-\xi_3)\gammaHone N^j(\delta\Phi^H_{ja}-ik_j\delta H_a)
\nonumber\\
&&\qquad\qquad
+\gammaHone\gammaHtwo N^j(\delta\Phi^H_{ja}-ik_j\delta H_a)
-\mu_1^2\xi_1 \delta F_a
\label{e:ThetaFlatCase},\qquad
\end{eqnarray} 
where $k^2=k^jk_j$, $\beta k = k_j N^j$, 
and contractions are done with the flat background
metric.  

This linearized gauge driver system,
Eqs.~(\ref{e:HFlatCase})--(\ref{e:ThetaFlatCase}), can be simplified
somewhat.  We note that Eq.~(\ref{e:HxFlatCase}) implies that
violations in the gauge constraint $\delta{\cal
C}^H_{ja}=\delta\Phi^H_{ja}- ik_j\delta H_a$ always decrease toward
zero exponentially on the timescale $1/\gammaHtwo$.  Since the system
is linear, we can (without loss of generality) confine our attention
to the constraint satisfying solutions, $\delta\Phi^H_{ja}=ik_j\delta
H_a$.  This condition and Eq.~(\ref{e:HFlatCase}) can be used to
eliminate the fields $\delta \Pi^H_a$ and $\delta\Phi^H_{ja}$ from the
system, resulting in the following simplified evolution system for
$\delta H_a$ and $\delta\theta_a$:
\begin{eqnarray}
&&\partial_t^2 \delta H_a 
+2\bigl[\mu_2(1-\xi_2)-i\beta k\bigr] \partial_t \delta H_a \nonumber\\
&&\quad
+\bigl[k^2(1-\beta^2)+\mu_1^2(1-\xi_1)
-2i\mu_2\beta k(1-\xi_2)\bigr] \delta H_a
\nonumber\\
&&\quad
=-\eta_1\delta\theta_a+\mu^2_1(1-\xi_1)\delta F_a,\label{e:NewHtFlatCase}
\\
&&\partial_t \delta \theta_a 
+\eta_1\delta \theta_a= 
-\bigl[2\mu_2(\xi_3-\xi_2)-i\beta k \bigr]\partial_t \delta H_a
\nonumber\\
&&\quad
-\Bigl[ k^2(1-\beta^2) -\mu_1^2\xi_1-2i\mu_2\beta k(1-\xi_2)\Bigr]
\delta H_{a}\nonumber\\
&&\quad-\mu_1^2\xi_1\delta F_a.\label{e:NewThetaFlatCase}
\end{eqnarray}

We note that the linearized gauge driver system,
Eqs.~(\ref{e:NewHtFlatCase})--(\ref{e:NewThetaFlatCase}), does not
depend on the metric perturbations $\delta\psi_{ab}$, except through
the target function $\delta F_a$.  This rather weak coupling means
that the gauge driver equations just respond to whatever target
$\delta F_a$ the gauge and spacetime geometry dictate.  It makes sense
then to investigate the intrinsic response of the gauge driver system
to a given $\delta F_a$.  We consider two simple test cases.  First,
in Sec.~\ref{s:Stability1} we consider the case where the target gauge
source function, $\delta F_a$, is time independent.  Second in
Sec.~\ref{s:Stability2}, we consider the more general case where
$\delta F_a=\delta F_a(t)$ is a prescribed function of time.  Then
finally, in Sec.~\ref{s:Stability3} we consider the more interesting
and realistic case where the gauge driver and GH Einstein systems are
coupled, using the target $\delta F_a$ appropriate for Bona-Mass\'o
slicing and a $\Gamma$-driver shift condition.

\subsection{Time Independent $\delta F_a$. }
\label{s:Stability1}

We consider first the case where the target gauge source function has
the form, $F_a=\delta F_a e^{ik_jx^j}$, for constant $\delta F_a$.  We
also assume that the shift of the background spacetime
vanishes: $\beta=0$.  In this case the general solution to
Eqs.~(\ref{e:NewHtFlatCase}) and (\ref{e:NewThetaFlatCase}) has the
form:
\begin{eqnarray}
&&\delta H_a(t) = \delta F_a + \sum_r\delta H_a^r e^{s_rt},\\ 
&&\delta\theta_a(t)=-\frac{k^2}{\eta_1}\delta F_a\nonumber\\
&&\qquad\quad-\sum_r\frac{k^2+2s_r\mu_2(\xi_3-\xi_2)-\mu_1^2\xi_1}
{s_r+\eta_1}\delta H_a^re^{s_rt},\qquad
\end{eqnarray}
where the $\delta H_a^r$ are constants and $s_r$ are the roots
(assumed to be non-degenerate) of the characteristic polynomial,
\begin{eqnarray}
&&0=  s_r^3 
+ \left[2\mu_2(1-\xi_2)+\eta_1\right]s_r^2\nonumber\\
&&\quad+ \left[k^2+\mu_1^2(1-\xi_1)+2\mu_2\eta_1(1-\xi_3)\right]
s_r+\eta_1\mu_1^2.\qquad
\label{e:CharacteristicPolynomial}
\end{eqnarray}
Equation~(\ref{e:CharacteristicPolynomial}) is the necessary and
sufficient condition that the solution satisfies
Eqs.~(\ref{e:NewHtFlatCase}) and (\ref{e:NewThetaFlatCase}).  The
three roots of Eq.~(\ref{e:CharacteristicPolynomial}) consist of a
real root, $s_0$, and a complex conjugate pair, $s_\pm$.

\begin{figure}
\centerline{\includegraphics[width=3in]{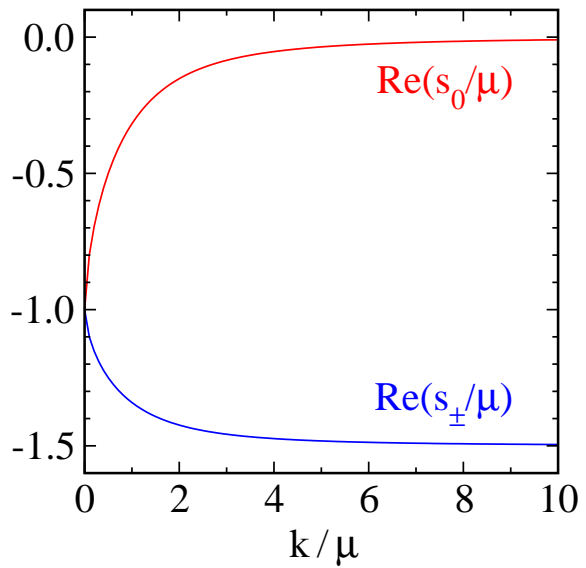}}
\caption{\label{f:FlatSpaceRoots}
Real part of the characteristic frequencies of the gauge driver
system: $s_0$ and $s_\pm$.}
\end{figure}

Figure~\ref{f:FlatSpaceRoots} illustrates the dependence of the real
parts of the roots, $s_0$ and $s_\pm$, on the wavenumber $k$ for the
case $\mu=\mu_1=\mu_2=\eta_1$ and $0=\xi_1=\xi_2=\xi_3$.  These roots
have strictly negative real parts for all $k$, so the gauge source
function $\delta H_a$ is always driven toward the target gauge source
function $\delta F_a$.  At least for this simple case, $\delta H_a$
approaches the target $\delta F_a$ exponentially.

Simple analytical expressions for the roots of the characteristic
polynomial, Eq.~(\ref{e:CharacteristicPolynomial}), exist in the
limits of small and large $k$.  The large $k$ limit is the most
interesting, because it describes the sufficently short wavelength
perturbations of any spacetime.  The asymptotic expressions for the
large $k$ roots are,
\begin{eqnarray}
{\mathrm{Re}}\left(s_0\right)&=&
-\eta_1\left(\frac{\mu_1}{k}\right)^2
+{\cal O}\left(k^{-4}\right),\\
{\mathrm{Re}}\left(s_{\pm}\right)
&=&-\mu_2(1-\xi_2)
- \frac{\eta_1}{2}+\frac{\eta_1}{2}\left(\frac{\mu_1}{k}\right)^{2}
+{\cal O}\left(k^{-4}\right).\nonumber\\
\end{eqnarray}
These results show that the $s_\pm$ modes are damped at approximately
the rate $\mu_2(1-\xi_2)+\eta_1/2$ in the large $k$ limit, while the
damping rate for the $s_0$ mode approaches zero.
These modes are stable for large enough $k$, then, as long
as $\eta_1\mu_1^2>0$ and $\mu_2(1-\xi_2)+\eta_1/2>0$.

\subsection{Time Dependent $\delta F_a$}
\label{s:Stability2}

\begin{figure}
\centerline{\includegraphics[width=3in]{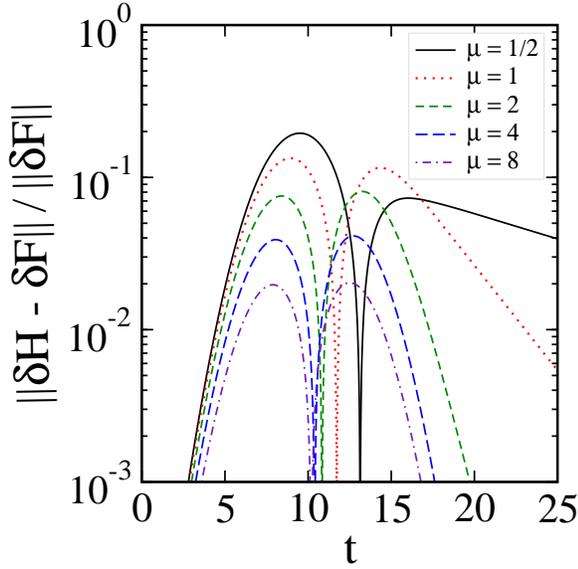}}
\caption{\label{f:FlatSpaceKVariationFt} Response of the gauge driver
system to a time dependent $\delta F_a$ of the form, $\delta F_a =
3+e^{-(t-10)^2/9}$, initial conditions $\delta H_a(o)=\delta F_a(0)$,
$\partial_t\delta H_a(0)=0$, and a range of values for the damping
parameter $\mu=\mu_1=\mu_2=\eta_1$. This test uses $k=1$ and
$\xi_1=\xi_2=\xi_3=0$.}
\end{figure}

Next we consider solutions to Eqs.~(\ref{e:NewHtFlatCase}) and
(\ref{e:NewThetaFlatCase}) for the case where $\delta F_a$ is a
specified function of time: $\delta F_a=\delta F_a(t)$.  In principle
the equations could be solved analytically by Laplace transforming the
equations in time, and solving for each frequency component of $\delta
H_a(t)$ separately.  Instead it is more straightforward, and perhaps
more instructive, to integrate the equations numerically for some
illustrative $\delta F_a(t)$.  We assume for this simple example that
the shift of the background spacetime vanishes, $\beta=0$, and the
other parameters that determine the system take the values:
$\mu=\mu_1=\mu_2=\eta_1$ and $0=\xi_1=\xi_2=\xi_3$.  We have solved
the resulting simplified equations numerically for the case $\delta
F_a(t)=3+e^{-(t-10)^2/9}$ with $k=1$.
Equations~(\ref{e:NewHtFlatCase}) and (\ref{e:NewThetaFlatCase})
require initial conditions for $\delta H_a$, $\partial_t \delta H_a$
and $\delta \theta_a$.  We use $\delta H_a(0)=\delta F_a(0)$,
$\partial_t \delta H_a(0)=0$ and $\mu\delta\theta_a(0)= -k^2\delta
H_a(0)$.  These initial data for $\delta H_a$ and its time derivative
were chosen to be fairly well matched with the target $\delta F_a$.
They are similar to the initial conditions used in our
more realistic tests in Sec.~\ref{s:NumericalTests}.  This target
$\delta F_a$ changes significantly for times near $t=10$, so this test
explores how well the gauge driver system is able to track an evolving
target $\delta F_a$.  Figure~\ref{f:FlatSpaceKVariationFt} shows that
the gauge driver equation is fairly successful (at the few percent
accuracy level) in driving $\delta H_a(t)$ toward $\delta F_a(t)$ for
$\mu\gtrsim 2$ even in this rather dynamical situation.

\subsection{Coupled Systems}
\label{s:Stability3}
Finally we investigate the stability of the coupled gauge driver and
GH Einstein equations for perturbations of flat spacetime.  The
perturbed Einstein system reduces to a relatively simple 
form\footnote{In this analysis we assume that the gauge constraint 
$\delta H_a + \delta \Gamma_a=0$ is satisfied.  The analysis of the
GH Einstein constraint evolution system in Ref.~\cite{Lindblom2006}
shows that violations of this constraint are damped exponentially
for perturbations of flat spacetime.}
in this case:
\begin{eqnarray}
\eta^{cd}\partial_c\partial_d \delta \psi_{ab} + \partial_a\delta H_b
+\partial_b \delta H_a = 0.\quad
\label{e:PerturbedEinstein}
\end{eqnarray}
We study the stability of the coupled system,
Eqs.~(\ref{e:NewHtFlatCase}), (\ref{e:NewThetaFlatCase}), and
(\ref{e:PerturbedEinstein}), by Laplace transforming the equations in
time, i.e., by considering solutions with time dependence $e^{st}$.
In this case Eqs.~(\ref{e:NewHtFlatCase}) and
(\ref{e:NewThetaFlatCase}) can be reduced to the single equation,
\begin{eqnarray}
\label{e:FLGauge}
P(s) \delta H_a
  &=& \mu^2_1\left(1-\frac{\xi_1 s}{s+\eta_1}\right)  \delta F_a,\qquad
\end{eqnarray}
where $P(s)$ is defined by, 
\begin{eqnarray}
P(s)&=&\hat s^2 
+ 2 \mu_2(1-\xi_2)\hat s + k^2 + \mu^2_1(1-\xi_1) \nonumber\\
&&- \frac{\eta_1}{s + \eta_1}\Bigl\{k^2-\mu^2_1\xi_1
-2i\beta k \mu_2(1-\xi_3)\nonumber\\
&&\qquad\qquad\,\,
+\hat s\bigl[2\mu_2(\xi_3-\xi_2)-i\beta k\bigr]\Bigr\}.\qquad
\label{e:Pdef}
\end{eqnarray}
We use the notation $\hat s \equiv s - i k \beta$.
The analogous
expressions for the Laplace transform of the GH Einstein system,
Eq.~(\ref{e:PerturbedEinstein}), are given by
\begin{eqnarray}
0&=&(\hat s^2+k^2)\delta \psi_{\hat t\hat t} - 2 \hat s \delta H_{\hat t},\\
0&=&(\hat s^2+k^2)\delta \psi_{\hat tj} 
- \hat s\delta H_j -i k_j\delta H_{\hat t},\\
0&=&(\hat s^2+k^2)\delta \psi_{jl} - i k_j\delta H_l - i k_l\delta H_j,
\end{eqnarray}
where $\delta H_{\hat t} = \delta H_t - N^j\delta H_j$, etc.  
These equations can be used to express $\delta H_a$ and $\delta\psi_{jl}$
in terms of $\delta \psi_{\hat ta}$ for the case $\hat s\neq 0$:
\begin{eqnarray}
\delta H_{\hat t} &=& \frac{\hat s^2+k^2}{2\hat s}\delta\psi_{\hat t\hat t},
\label{e:PerturbedHt}\\
\delta H_j &=&\frac{\hat s^2+k^2}{\hat s^2}
\left(\hat s\,\delta\psi_{\hat tj}-\frac{1}{2}
ik_j\delta\psi_{\hat t\hat t}\right),\label{e:PerturbedHi}\\
\delta\psi_{jl}&=&\hat s^{-2}\left(i\hat sk_j\delta\psi_{\hat tl}+
i\hat sk_l\delta\psi_{\hat tj}+k_jk_l\delta\psi_{\hat t\hat t}\right).
\label{e:PerturbedPsiIJ}
\end{eqnarray}
The case $\hat s=0$ case is essentially trivial: In this case
$k^2\delta\psi_{\hat t\hat t}=0$, $k^2\psi_{\hat tj}=ik_j\delta H_{\hat t}$,
$k^2\delta\psi_{jl}=ik_j\delta H_l+ik_l\delta H_j$ and $\delta
H_a=\delta F_a$.  The metric perturbation in this case is pure gauge
(an infinitesimal coordinate transformation generated by the time
independent $\delta H_a/k^2$), and the gauge source function $\delta
H_a$ is identical to the target $\delta F_a$ in this case.  So we
focus on the $\hat s\neq 0$ case for the remainder of this discussion.

We consider in detail now the coupled gauge driver system for the case
of Bona-Mass\'o slicing and the $\Gamma$-driver shift condition.  The
perturbed flat space limit of $\delta F_a$ for the Bona-Mass\'o driver,
Eq.~(\ref{e:OnePlusLogF1}), is given by
\begin{eqnarray}
\delta F_{\hat t} &=&\left[\hat s\,\frac{f(1)-\rho_1}{2f(1)} 
- \frac{i\beta k\rho_1}{2f(1)}\right]
\delta\psi_{\hat t\hat t}\nonumber\\
&&+i(\rho_1-1)k^l\delta\psi_{\hat t l}
-\frac{1}{2}(\rho_1-1)\hat s \delta \psi^l{}_l,
\label{e:PerturbedBM}
\end{eqnarray}
while the target for the $\Gamma$-driver shift condition,
Eq.~(\ref{e:NewGammaDriver}), reduces to
\begin{eqnarray}
\delta F_j &=& 
(\hat s - \rho_2 s)\delta \psi_{\hat t j}
+i\left(\frac{\nu\rho_2 s}{s+\eta_2}
-1\right)\delta\psi_{jl}k^l 
\nonumber\\
&&
-\frac{i}{2}k_j\delta\psi_{\hat t\hat t}
-\frac{i}{2}\left[\frac{\nu\rho_2s( 1+\lambda)}
{s+\eta_2}-1\right]
k_j\delta \psi{}^l{}_l
.\qquad\label{e:PerturbedGammaDriver}
\end{eqnarray}
The spatial metric perturbations, $\delta\psi_{jl}$, that appear in
Eqs.~(\ref{e:PerturbedBM}) and (\ref{e:PerturbedGammaDriver}) can be
replaced by $\delta\psi_{\hat ta}$ using Eq.~(\ref{e:PerturbedPsiIJ}):
\begin{eqnarray}
\delta F_{\hat t}&=&\left[\hat s\,\frac{f(1)-\rho_1}{2f(1)} 
- \frac{i\beta k\rho_1}{2f(1)}-\frac{k^2(\rho_1-1)}{2\hat s}\right]
\delta\psi_{\hat t\hat t},\qquad \label{e:PerturbedBM1}\\
\delta F_j&=&\frac{i}{2}\left[
\frac{k^2}{\hat s^2}\left(\nu \rho_2 s 
\frac{1-\lambda}{s+\eta_2}-1\right)-1\right]
k_j\delta\psi_{\hat t\hat t}\nonumber\\
&&-\left[\frac{k^2}{\hat s}\left(\frac{\nu \rho_2 s}{s+\eta_2}-1\right)
+\rho_2 s - \hat s\right]\delta\psi_{\hat tj}\nonumber\\
&&+ \frac{\nu \rho_2 s\lambda}{\hat s(s+\eta_2)} k_j k^l\delta\psi_{\hat tl}
.\qquad\label{e:PerturbedGammaDriver1}
\end{eqnarray}

Now substitute these expressions for $\delta F_a$,
Eqs.~(\ref{e:PerturbedBM1}) and (\ref{e:PerturbedGammaDriver1}), and
the expressions for $\delta H_a$, Eqs.~(\ref{e:PerturbedHt}) and
(\ref{e:PerturbedHi}), into the perturbed gauge driver
Eq.~(\ref{e:FLGauge}).  The result is a system of linear algebraic
equations for $\delta\psi_{\hat ta}$.  This system can be decoupled and
non-trivial solutions exist if and only if the frequency $s$
satisfies one of the following characteristic polynomials:
\begin{eqnarray} 
0&=&\frac{\hat s^2+k^2}{\hat s}P(s)
+\mu^2_1\biggl(1-\frac{\xi_1 s}{s+\eta_1}\biggr)\times\nonumber\\
&&\qquad\,\,
\biggl[\hat s \frac{\rho_1-f(1)}{f(1)}
+\frac{i\beta k \rho_1}{f(1)}
+(\rho_1-1)\frac{k^2}{\hat s}\biggr],
\label{e:TimeMode}\qquad\\
0&=&\frac{\hat s^2+k^2}{\hat s}P(s)
+\mu^2_1\biggl(1-\frac{\xi_1 s}{s+\eta_1}\biggr)\times\nonumber\\
&&\qquad\quad\,\biggl[
\frac{  k^2}{\hat s} \left(\nu \rho_2 s\frac{1-\lambda}{s+\eta_2}-1\right)
+\rho_2 s -\hat s\biggr],
\label{e:LongitudinalMode}\\
0&=&
\frac{\hat s^2+k^2}{\hat s}P(s)
\nonumber\\
&&\qquad\quad\,+\mu^2_1 
\biggl[\frac{ k^2}{\hat s} 
\left(\frac{\nu \rho_2 s}{s+\eta_2}-1\right)
+\rho_2 s -\hat s\biggr],
\label{e:TransverseMode}
\end{eqnarray} 
where $P(s)$ is defined in Eq.~(\ref{e:Pdef}).

The flat space stability analysis presented here is relevant to
generic spacetimes when the wavenumber $k$ of the
perturbation becomes sufficiently large.  We have solved the characteristic
polynomials in Eqs.~(\ref{e:TimeMode})--(\ref{e:TransverseMode})
in this limit.  The leading order expressions for the real parts
of these roots are given as follows.  For the time slicing modes
(in which $\delta \psi_{\hat t\hat t}\neq0$),
the roots of Eq.~(\ref{e:TimeMode}), we have
\begin{eqnarray}
\mathrm{Re}\left(s\right)&=& -\frac{\eta_1\mu^2_1\rho_1}{(1-\beta^2)^2k^2}
+{\cal O}(k^{-4}),\\
\mathrm{Re}\left(s\right)&=&
\pm\frac{1}{4}\biggl\{\bigl[\eta_1+2\mu_2(1-\xi_2)\bigr]^2
\nonumber\\
&&\qquad-4\mu_1^2\rho_1(1-\xi_1)\frac{1-f(1)
\pm\beta}{f(1)}\biggr\}^{1/2}\nonumber\\
&&-\frac{1}{4}\bigl[\eta_1+2\mu_2(1-\xi_2)\bigr]
+{\cal O}\left(k^{-2}\right).\label{e:TimeMode2}
\end{eqnarray}
The asymptotic forms of the roots of the longitudinal modes
(in which $k^j\delta \psi_{\hat t j}\neq0$),
Eq.~(\ref{e:LongitudinalMode}), are given by
\begin{eqnarray}
\mathrm{Re}\left(s\right)&=& -\eta_2
+{\cal O}(k^{-2}),\\
\mathrm{Re}\left(s\right)&=&
\pm\frac{1}{4}\Bigl\{\bigl[\eta_1+2\mu_2(1-\xi_2)\bigr]^2
\nonumber\\
&&\qquad-4\mu_1^2\rho_2(1-\xi_1)[1\pm\beta- \nu(1-\lambda)]\Bigr\}^{1/2}
\nonumber\\&&-\frac{1}{4}\bigl[\eta_1+2\mu_2(1-\xi_2)\bigr]
+{\cal O}(k^{-2}).\label{e:LongitudinalMode2}
\end{eqnarray}
Finally the asymptotic forms of the roots of the transverse modes
(in which $[k^2g^{ij}-k^ik^j]\delta\psi_{\hat t j}\neq0$),
Eq.~(\ref{e:TransverseMode}), are given by
\begin{eqnarray}
\mathrm{Re}\left(s\right)&=& -\eta_2
+{\cal O}(k^{-2}),\\
\mathrm{Re}\left(s\right)&=&
\pm\frac{1}{4}\Bigl\{\bigl[\eta_1+2\mu_2(1-\xi_2)\bigr]^2
\nonumber\\
&&\qquad-4\mu_1^2\rho_2(1-\xi_1)[1\pm\beta- \nu]\Bigr\}^{1/2}
\nonumber\\
&&-\frac{1}{4}\bigl[\eta_1+2\mu_2(1-\xi_2)\bigr]+{\cal O}(k^{-2}).
\label{e:TransverseMode2}
\end{eqnarray}
All four $\pm$ sign combinations represent distinct roots in
Eqs.~(\ref{e:TimeMode2}), (\ref{e:LongitudinalMode2}), and
(\ref{e:TransverseMode2}).  Stability of the gauge driver system
requires $\mathrm{Re}(s)< 0$.  Therefore, stability of the short
wavelength modes requires the following inequalities on the system
parameters:
\begin{eqnarray}
0&<&\eta_1\rho_1,\label{e:Inequality1}\\
0&<&\eta_1+2\mu_2(1-\xi_2),\\
0&<&\rho_1(1-\xi_1)\frac{1-f(1)\pm\beta}{f(1)},\\
0&<&\eta_2,\\
0&<&\rho_2(1-\xi_1)\bigl[1\pm\beta -\nu(1-\lambda)\bigr],\\
0&<&\rho_2(1-\xi_1)\bigl[1\pm\beta -\nu\bigr].\label{e:Inequality6}
\end{eqnarray}
We note thate these conditions can be satisfied for small values of
$\beta$ by taking $\eta_1>0$, $\eta_2>0$, $\mu_2>0$, $\xi_1<1$,
$\xi_2< 1$, $\rho_1>0$, $\rho_2>0$, $0<f(1)<1$, $\nu<1$,
$\nu(1-\lambda)<1$.

%
\begin{figure}
\centerline{\includegraphics[width=3in]{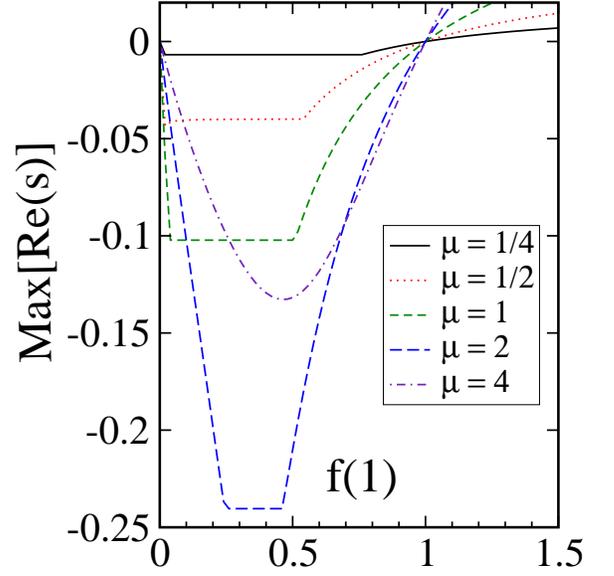}}
\caption{\label{f:MaxS_f_VariousMu} Maximum damping rate of the
modes as a function of the Bona-Mass\'o slicing condition parameter
$f(1)$.  The other parameters used for this case are
$k=1$, $\beta=0$, 
$\mu=\mu_1=\mu_2=\eta_1=\frac{1}{32}\eta_2$, $\xi_1=\xi_2=\xi_3=0$,
$\rho_1=\rho_2=\frac{1}{2}$, $\nu=\frac{3}{4}$, and
$\lambda=-\frac{1}{3}$.  }
\end{figure}

We have also explored the roots of these characteristic polynomials
numerically.  Figure~\ref{f:MaxS_f_VariousMu} illustrates
$\mathrm{Max[Re(}s)]$, the root of these equations having the largest
real part, as a function of the parameter $f(1)$ that characterizes
the Bona-Mass\'o slicing condition in this flat space limit.  The
curves correspond to the roots for the driver system with various
values of $\mu=\mu_1=\mu_2=\eta_1=\frac{1}{32}\eta_2$,
$\lambda=-\frac{1}{3}$, $\nu=\frac{3}{4}$, and $k=1$.  These parameter
values were chosen because they satisfy the inequalities in
Eqs.~(\ref{e:Inequality1})--(\ref{e:Inequality6}), and because they
perform fairly well for the 3D numerical tests discussed in
Sec.~\ref{s:NumericalTests}.  We see that the maximum real part of $s$
is negative for $f(1)$ in the range $0<f(1)<1$, and so the coupled
gauge driver system is stable for these values.  The system is most
stable for $f(1)\approx\frac{1}{2}$, so we adopt this value in our
numerical tests of the gauge driver system in
Sec.~\ref{s:NumericalTests}.  We also note that the standard value,
$f(1)=2$, used for one-plus-log slicing by most of the numerical
relativity community~\cite{Balakrishna1996, Alcubierre2002,
Campanelli2006a, Baker2006a} is unstable when used in our gauge driver
equations.  This does not imply that $f(1)=2$ is a bad choice when
used in a standard three-plus-one evolution, only that it is unstable
when used with our gauge driver system.

\begin{figure}
\centerline{\includegraphics[width=3in]{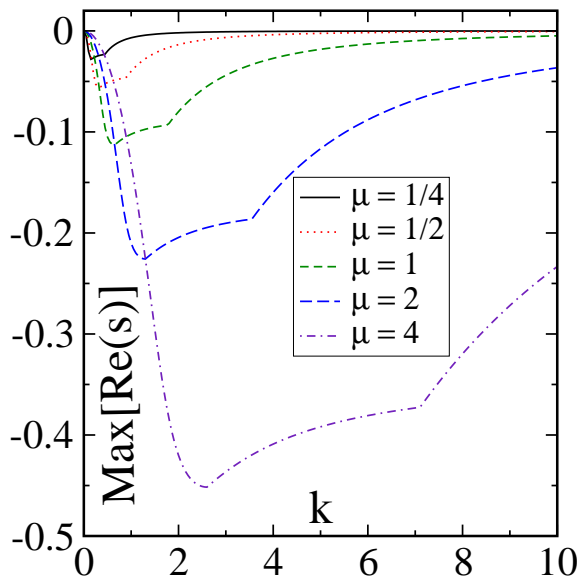}}
\caption{\label{f:MaxS_k_VariousMu} Maximum damping rate of the modes
as a function of the wavenumber $k$.  The other parameters used
for this case are $\beta=0$, $\mu=\mu_1=\mu_2=\eta_1=\frac{1}{32}\eta_2$,
$\xi_1=\xi_2=\xi_3=0$, $\rho_1=\rho_2=\frac{1}{2}$,
$f(1)=\frac{1}{2}$, $\nu=\frac{3}{4}$, and $\lambda=-\frac{1}{3}$.  }
\end{figure}

Figure~\ref{f:MaxS_k_VariousMu} illustrates the $k$ dependence of
$\mathrm{Max[Re(}s)]$ for a range of values of the gauge driver
damping coefficients $\mu=\mu_1=\mu_2=\eta_1=\frac{1}{32}\eta_2$.  For short
wavelength perturbations, i.e. for values of $k$ with  $k\gtrsim\mu$,
$\mathrm{Max[Re}(s)]$ decreases as $\mu$ increases.  Thus the
solutions with large $k$ are damped more effectively as $\mu$
increases.  However, for long wavelength perturbations, i.e. for
values of $k$ with $k\lesssim\mu$, $\mathrm{Max[Re}(s)]$ increases as $\mu$
increases.  Thus the solutions with small $k$ are less efficiently
damped as $\mu$ increases.  It follows that there is an optimal value
of $\mu$ to use for any particular problem: choose $\mu\approx k_c$,
where $1/k_c$ corresponds to the lengthscale on which the gauge condition
needs to be enforced most effectively.
%

\begin{figure}
\centerline{\includegraphics[width=3in]{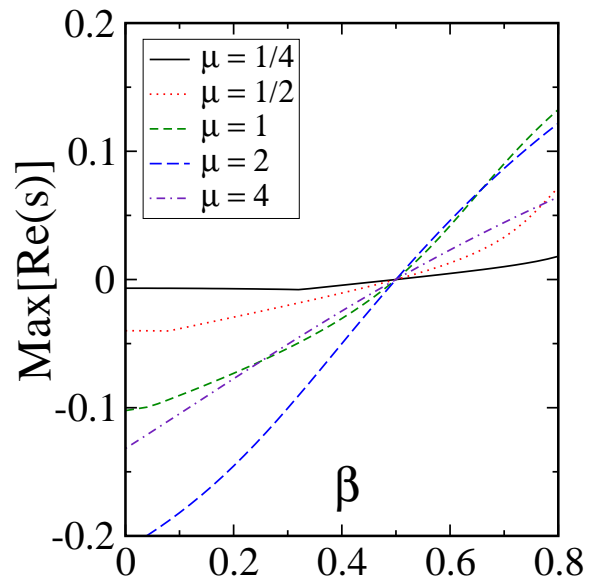}}
\caption{\label{f:MaxS_Beta_VariousMu} Maximum damping rate of the
modes as a function of the shift parameter $\beta$.  The other
parameters used for this case are $k=1$,
$\mu=\mu_1=\mu_2=\eta_1=\frac{1}{32}\eta_2$, $\xi_1=\xi_2=\xi_3=0$,
$\rho_1=\rho_2=\frac{1}{2}$, $f(1)=\frac{1}{2}$, $\nu=\frac{3}{4}$,
and $\lambda=-\frac{1}{3}$.  }
\end{figure}

Figure~\ref{f:MaxS_Beta_VariousMu} illustrates the dependence of
$\mathrm{Max[Re(}s)]$ on the background shift parameter
$\beta$ for a range of values of the gauge driver
damping coefficients $\mu=\mu_1=\mu_2=\eta_1=\frac{1}{32}\eta_2$.  For
small values of $\beta$ we see that the system is stable, however for
$\beta>\frac{1}{2}$ the system becomes unstable.  This instability may
be important in more realistic problems that involve black holes. Even
for single black-hole spacetimes, the usual time-independent
coordinate representations have non-vanishing shifts with
$\beta\approx 1$ near the horizon.  Binary black hole spacetimes also
use large shifts (with $\beta>1$ in many cases) when coordinates that
co-rotate with the black holes are used.  We explore the stability of
this gauge-driver system for the case of single black-hole spacetimes
in Sec.~\ref{s:NumericalTests}.
%

We have also examined several other slicing and shift conditions using
these perturbation techniques.  As a consequence of section
\ref{s:Stability1}, our gauge driver system is stable for harmonic
slicing $\delta F_{\hat t}=0$ and harmonic shift $\delta F_i=0$.  We
also find that the combinations of a stable Bona-Mass\'o slicing
condition with harmonic shift, and of harmonic slicing with the
$\Gamma$-driver shift condition are stable.  However, we find that the
maximal slicing and $\Gamma$-freezing conditions are unconditionally
unstable when enforced through our gauge driver equations.
%

\section{Numerical Tests}
\label{s:NumericalTests}

In this section we describe the results of 3D numerical tests of the
gauge driver system.  We consider two cases: first a Schwarzschild
black hole with perturbed lapse and shift, and second a Schwarzschild
black hole with a superimposed outgoing physical gravitational wave
pulse.  The full coupled non-linear GH Einstein and gauge driver
systems are solved numerically for these cases.  We measure the
stability and effectiveness of the gauge driver system in these tests
as it attempts to drive the gauge toward Bona-Mass\'o slicing
and $\Gamma$-driver shift conditions.

These numerical tests are conducted using the infra\-structure of the
Caltech/Cornell Spectral Einstein Code (SpEC).  This code uses
pseudospectral collocation methods, as described for example in
Refs.~\cite{Kidder2005,Boyle2006}.  We use the generalized harmonic
form of the Einstein equations, as described in
Ref.~\cite{Lindblom2006}.  The evolution equations for the combined GH
Einstein and the gauge driver system are integrated in time using the
adaptive fifth-order Cash-Karp method~\cite{CashKarp1990}.  We
use a form of spectral filtering, as described in
Ref.~\cite{Kidder2005}, that sets to zero in each time step the
changes in the top four \emph{tensor} spherical harmonic expansion
coefficients of each of our evolved quantities.  This filtering step
is needed to eliminate an instability associated with the inconsistent
mixing of tensor spherical harmonics in our approach.

Initial conditions are needed for any evolution of the combined GH
Einstein and gauge driver systems, and these initial data consist of
the spacetime metric $\psi_{ab}$, the gauge source function $H_a$, and
their time derivatives.  For the tests described here we take the
initial spacetime metric $\psi_{ab}$ to be the Schwarzschild geometry
plus perturbations as described in Secs.~\ref{s:BHGaugePerturbations} and
\ref{s:BHPhysicalPerturbations}.  We set the time derivatives of the
spatial components of the metric to zero for the pure gauge
perturbation test in Sec.~\ref{s:BHGaugePerturbations}, and equal to
the appropriate time derivative of the superimposed physical
gravitational wave pulse for the test in
Sec.~\ref{s:BHPhysicalPerturbations}. The remaining initial data
needed for these evolutions, $\partial_t N$, $\partial_t N^i$, $H_a$,
and $\partial_t H_a$, are pure gauge quantites.  The time derivatives
of the lapse and shift are chosen here to ensure that $H_a$ satisfies
the desired gauge condition, $H_a=F_a$, initially.  And finally the
initial value of $H_a$ is chosen here to ensure that the gauge
constraint, ${\cal C}_a=H_a+\Gamma_a=0$, vanishes initially.

\subsection{Black Hole with Gauge Perturbation}
\label{s:BHGaugePerturbations}

For this test we consider a Schwarzschild black hole with
perturbations in the lapse and shift.  For the unperturbed hole we use
isotropic spatial coordinates and maximal time
slices~\cite{Estabrook1973,Cook2004}.  The unperturbed spatial metric
in this representation is given by,
\begin{eqnarray}
ds^2&=&g_{ij}dx^idx^j
=\left(\frac{R}{r}\right)^2\left(dx^2
+ dy^2 + dz^2\right),\label{e:IsotropicMetric}\qquad
\end{eqnarray}
where $r^2=x^2+y^2+z^2$, and $R(r)$ (the areal radius)
satisfies the differential equation,
\begin{eqnarray}
\frac{{\mathrm d}R}{{\mathrm d}r} &=&
\frac{R}{r}\sqrt{1-\frac{2M}{R}+\frac{C^2}{R^4}}.
\end{eqnarray}
The constant $M$ is the mass of the hole, and $C$ is a parameter that
specifies the particular maximal slicing.  Finally, the unperturbed
lapse $N$ and shift $N^i$ for this representation of Schwarzschild are
given by,
\begin{eqnarray}
N&=& \sqrt{1-\frac{2M}{R}+\frac{C^2}{R^4}},\label{e:SchwarzschildLapse}\\
N^i&=&\frac{C\hat r^i}{R^2}
\left(1-\frac{2M}{R}+\frac{C^2}{R^4}\right),\label{e:SchwarzschildShift}
\end{eqnarray}
where $\hat r^i$ is the outward directed radial unit vector:
$g_{ij}\hat r^i \hat r^j=1$.

We perturb this spacetime by changing the initial values of the lapse
and shift, and their time derivatives.  This type of perturbation
changes the spacetime coordinates (or gauge) of the solution, but not
its geometry.  For these tests we perturb the lapse and shift of
Eqs.~(\ref{e:SchwarzschildLapse}) and (\ref{e:SchwarzschildShift}) by
adding,
\begin{eqnarray}
\delta N &=& A \sin(2\pi r/r_0)e^{-(r-r_c)^2/w^2}Y_{lm},
\label{e:PerturbedSchwarzschildLapse}\\
\delta N^i &=& A \sin(2\pi r/r_0)e^{-(r-r_c)^2/w^2} Y_{lm}\hat r^i,
\label{e:PerturbedSchwarzschildShift}
\end{eqnarray}
where $Y_{lm}$ is the standard scalar spherical harmonic.  In our
numerical tests we use the background metric with $C=1.73M^2$, and
perturbations with $A=0.01$, $r_c=15M$, $w=3M$, $l=2$, $m=0$, and
various values of the radial wavelength $r_0$.

We perform these numerical tests on a computational domain consisting
of a spherical shell that extends from $r=0.78M$ (just inside the
horizon in these coordinates) to $r=30M$.  We divide this domain into
eight subdomains.  In each subdomain we express each Cartesian
component of each dynamical field as a sum of Chebychev polynomials of
$r$ (through order $N_r-1$) multiplied by scalar spherical harmonics
(through order $L$).  The radii of the inner and outer edges of the
various subdomains are adjusted to distribute the truncation error for
this problem more or less uniformly.  
The specific radii of the
subdomain boundaries used in this test are $0.78M,$ $2.38M,$ $4.6M,$
$8.83M,$ $13.07M,$ $17.30M,$ $21.53M,$ $25.77M,$ and $30.0M$
respectively.  The values of the parameters associated with the gauge
driver system used for this test are: $\nu=\frac{3}{4}$,
$\lambda=-\frac{1}{3}$, $\rho_1=\rho_2=\frac{1}{2}$,
$\xi_1=\xi_2=\xi_3=0$, and various values of the parameter
$\mu=\mu_1=\mu_2=\eta_1=\frac{1}{32}\eta_2$.  The Bona-Mass\'o slicing
condition includes a target value for the extrinsic curvature $K_0$;
for this test we set $K_0=0$.

\begin{figure}[t]
\vspace{0.1cm}
\centerline{
\includegraphics[width=3in]{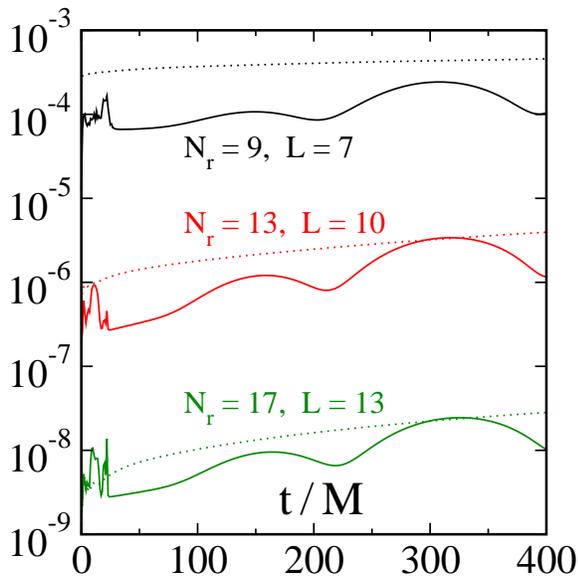}}
\caption{\label{f:BHGaugeConstraints}Solid curves show the constraints
of the GH Einstein system $||\,{\cal C}_{\mathrm{GH}}||$, dotted curves
show the constraints of the gauge driver system $||\,{\cal C}_{\mathrm
H}||$ for a test with $\mu=1.0/M$ and radial wavelength $r_0=6.0M$ 
evolved at several numerical resolutions.}
\end{figure}

Figure~\ref{f:BHGaugeConstraints} illustrates the constraint
violations for a set of representative evolutions from this test, and
demonstrates the exponential convergence of our numerical method.  The
solid curves represent the constraints associated with the GH Einstein
system, while the dotted curves represent the constraints of the gauge
driver system.  We measure the constraint violations of the GH
Einstein system for these tests using the norm $||\,{\cal
C}_{{\mathrm{GH}}}||$ defined in Eq.~(71) of Ref.~\cite{Lindblom2006}.
The norm $||\,{\cal C}_{{\mathrm{GH}}}||$ is scaled so that it becomes
of order unity when constraint violations start to dominante the
solution.  We define an analogous norm $||\,{\cal C}_{{\mathrm{H}}}||$
for the gauge driver system:
\begin{eqnarray}
||\,{\cal C}_{\mathrm H}||^2&=&
\int\!\sqrt{g}\, m^{ab}g^{ij}\left({\cal C}^H_{ia}{\cal C}^H_{jb}
+g^{kl}{\cal C}^H_{ika}{\cal C}^H_{jlb}\right)d^{\,3}x\nonumber\\
&&\times\biggl[\int\!\sqrt{g}\, m^{cd}g^{ij}\Bigl(\partial_iH_c\partial_jH_d
+\partial_i\Pi^H_c\partial_j\Pi^H_d\nonumber\\
&&\qquad\qquad\qquad +g^{kl}\partial_i\Phi^H_{kc}\partial_j\Phi^H_{ld}
\Bigr)d^{\,3}x\biggr]^{-1}.\qquad
\end{eqnarray}
The quantity $m^{ab}$ is a positive definite matrix, which we set to
the identity matrix, $m^{ab}=\delta^{ab}$, for these tests.
Figure~\ref{f:BHGaugeConstraints} shows the constraints for a
particular test run with $\mu=1.0/M$ and $r_0=6.0M$.  The analogous
graphs for the other tests reported here are qualitatively similar,
with somewhat larger but still convergent ``spikes'' 
in $||\,{\cal C}_{\mathrm GH}||$ at early times ($t\lesssim 25M$) 
for the $\mu=0.5/M$ case.
Figure~\ref{f:BHGaugeConstraints} shows that the constraints are well
satisfied in our evolutions, and demonstrates that our numerical
methods are (exponentially) convergent.  The mild power law growth in
the constraints seen at late times is sublinear,
and is not something that concerns us.
%

\begin{figure}[t]
\vspace{0.1cm}
\centerline{
\includegraphics[width=3in]{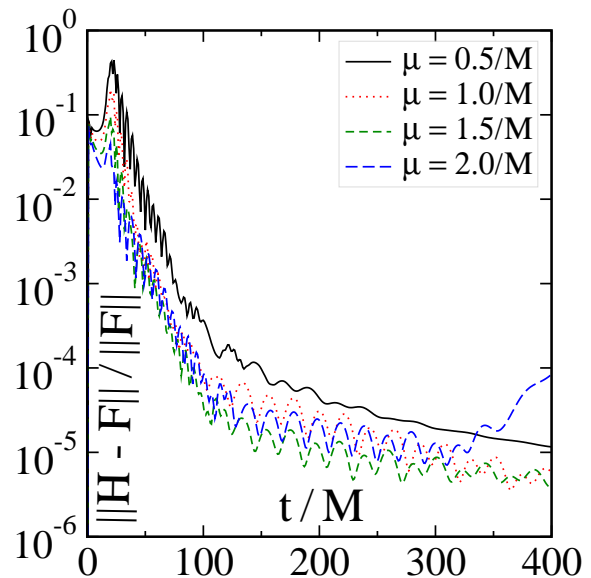}}
\caption{\label{f:HMinusFVaryingMu}Effectiveness of the gauge driver
equation is demonstrated by showing $||H-F||/||F||$ for evolutions
with radial wavelength $r_0=6M$ and several values of the gauge damping 
parameter $\mu M\in\{\frac{1}{2},1,\frac{3}{2},2\}$.  These tests evolve a
Schwarzschild black hole with strongly perturbed lapse and shift. }
\end{figure}
%

Figure~\ref{f:HMinusFVaryingMu} illustrates the effectiveness of the
gauge driver system, at least for this test problem.  We measure the
difference between the gauge source function $H_a$ and the target
function to which it is being driven, $F_a$, using the following
$L^2$ norm:
\begin{eqnarray}
\frac{||H-F||^2}{||F||^2}&=&
\frac{\int\!\sqrt{g}\,m^{ab}(H_a-F_a)(H_b-F_b)\,d^{\,3}x}
{\int\!\sqrt{g}\, m^{cd}F_cF_d\,d^{\,3}x},\qquad
\end{eqnarray}
where (as before) the matrix $m^{ab}$ is set to the identity,
$m^{ab}=\delta^{ab}$, for these tests.  This norm is scaled so that
$H_a$ bears little resemblance to the target $F_a$ whenever the norm
becomes of order unity.  Figure~\ref{f:HMinusFVaryingMu} shows that
the gauge perturbation used in this test violates the desired gauge
conditions rather severly at early times.  The norm $||H-F||/||F||$ is
driven to values as large as 0.7 at about $t=20M$ when the ingoing
part of the gauge perturbation interacts most strongly with the black
hole.  After this initial interaction, the gauge driver system takes
over and effectively drives $||H-F||/||F||$ to values below $10^{-3}$
on timescales of $40M$ to $60M$, depending on the value of the gauge
damping parameter $\mu$ used in the evolution.
Figure~\ref{f:HMinusFVaryingMu} shows evolutions of gauge
perturbations with radial wavelength $r_0=6M$, and several values of
the damping parameter $\mu M\in\{\frac{1}{2},1,\frac{3}{2},2\}$ computed
with numerical resolution $N_r=17$ and $L=13$.  The gauge driver
system is more effective at reducing $||H-F||/||F||$ quickly at early
times, $t<75M$, for larger values of $\mu$.

\begin{figure}[t]
\vspace{0.1cm}
\centerline{
\includegraphics[width=3in]{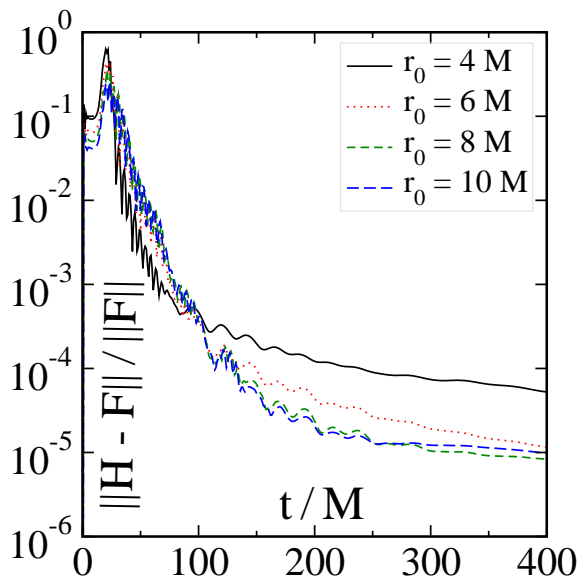}}
\caption{\label{f:HMinusFVaryingWavelength}Effectiveness of the gauge driver
equation is demonstrated by showing $||H-F||/||F||$ for evolutions
with $\mu=0.5/M$ and several values of the radial wavelength of the
perturbation $r_0\in\{4M,6M,8M,10M\}$.  These tests evolve a
Schwarzschild black hole with strongly perturbed lapse and shift.}
\end{figure}

The $\mu=2/M$ case shown in Fig.~\ref{f:HMinusFVaryingMu} has a mild
instability, that first appears at about $t=300M$.  This is a gauge
instability since it does not affect any of the constraint quantities.
Larger values of $\mu$ are progressively more unstable.  This
instability may be related to the rather unusual dispersion relation
for this gauge driver, as shown for the flat space case in
Fig.~\ref{f:MaxS_k_VariousMu}.  The gauge driver equation becomes
increasingly ineffective for driving the long wavelength components of
$H_a$ toward $F_a$ as $\mu$ increases.  
This poor damping efficiency
for long wavelengths, together with our rather simplistic boundary
conditions or the inherent instability associated with
large shifts, may well be the cause of this instability.
Figure~\ref{f:HMinusFVaryingWavelength} provides some additional
insight into the way the gauge driver equation responds to different
perturbations.  The evolutions shown in
Fig.~\ref{f:HMinusFVaryingWavelength} are all performed with
$\mu=0.5/M$ but several different values of the radial
wavelength of the gauge perturbation: $r_0\in \{4M,6M,8M,10M\}$.  These
tests show that the gauge driver system causes $||H-F||/||F||$ to
approach zero more quickly (at least at early times) 
for shorter wavelength perturbations.  
Our hope is that this ability to efficiently control short wavelength
features of the gauge is what will be needed to prevent the kinds of
localized gauge singularities that often appear in our evolutions of
binary black hole spacetimes.
%

\subsection{Black Hole with Physical Perturbation}
\label{s:BHPhysicalPerturbations}

Our second numerical test of the gauge driver system uses a
Schwarzschild black hole with a superimposed outgoing gravitational
wave pulse, as described in Refs.~\cite{Kidder2005,Rinne2007}. The
background solution is a Schwarzschild black hole in Kerr-Schild
coordinates,
\begin{equation}
  d s^2 = -d t^2 + \frac{2M}{r} (d t +
  d r)^2 + d x^2 + d y^2 + d z^2,
\end{equation}
where $r^2=x^2+y^2+z^2$ and $M$ is the mass.  We superimpose an
odd-parity outgoing quadrupolar gravitational wave perturbation
constructed using Teukolsky's method \cite{Teukolsky1982}. Its
generating function is taken to be a Gaussian, $G(r) = A \exp [-(r -
r_c)^2/w^2]$, with $A = 4 \times 10^{-3}$, $r_c = 5M$, and $w = 1.5 M$.
Using this perturbed Schwarzschild solution as the input conformal
metric, the full non-linear initial value equations (in the conformal
thin sandwich formulation) are solved to obtain initial data that
satisfy the constraints~\cite{Pfeiffer2004}.  This procedure yields
initial values for the spatial metric, extrinsic curvature, lapse, and
shift.  We note that the resulting solution to the constraints is
still nearly (but not completely) outgoing.
 
The computational domain for this test problem is taken to be a
spherical shell extending from $r = 1.9 M$ (just inside the horizon in
these coordinates) out to $r = 41.9M$. This domain is subdivided into
four spherical-shell subdomains of width $\Delta r = 10 M$.  On each
subdomain, the numerical solution is expanded in Chebyshev polynomials
and spherical harmonics as before.
For these tests we use
numerical resolutions with $N_r \in \{ 21, 31, 41, 51 \}$ coefficients
per subdomain for the Chebyshev series and $l \leqslant L$ with $L \in
\{ 8, 10, 12, 14 \}$ for the spherical harmonics.

\begin{figure}[t]
\vspace{0.1cm}
\centerline{
\includegraphics[width=3in]{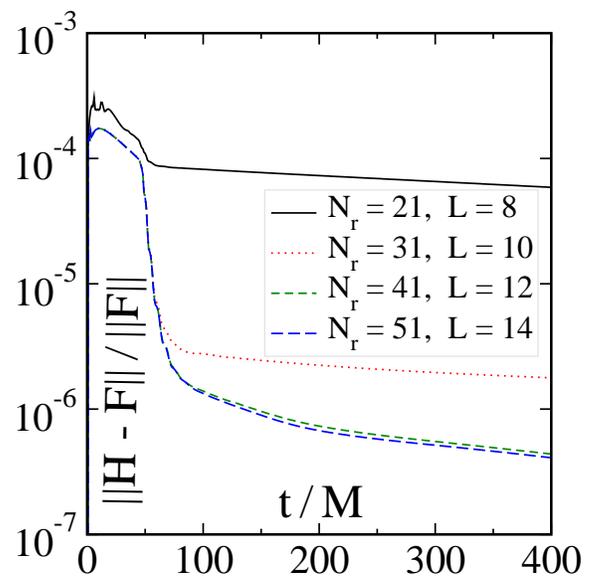}}
\caption{\label{f:HMinusFPhysicalPerturbation}Effectiveness of the
gauge driver equation is demonstrated by showing $||H-F||/||F||$ for
evolutions with $\mu=0.25/M$ obtained with a variety of numerical
resolutions.  This test uses a Schwarzschild black hole with a
superimposed outgoing gravitational wave pulse.}
\end{figure}

Figure~\ref{f:HMinusFPhysicalPerturbation} illustrates the
effectiveness of the gauge driver equation for imposing the
Bona-Mass\'o slicing and $\Gamma$-driver shift conditions in
evolutions of a Schwarzschild black hole with physical gravitational
wave perturbation.  These tests were performed with the gauge damping
parameter $\mu=0.25/M$.  
For this test we set the target value for the extrinsic curvature 
$K_0$ to that of an unperturbed Kerr-Schild spacetime.
The various curves in Fig.~\ref{f:HMinusFPhysicalPerturbation}
illustrate how $||H-F||/||F||$ changes for evolutions performed with
different numerical resolutions.  The results are qualitatively
similar to those of the first test: the black hole with physical
gravitational wave perturbation does not satisfy the target gauge
conditions exactly at early times, but the gauge driver equation
reduces $||H-F||/||F||$ to very small values by about $t=75M$.  This
test is less severe in some sense than our first pure gauge
perturbation test, since the initial data in this case contains an
outgoing gravitational wave pulse that never interacts very strongly
with the black hole.
%

\section{Discussion}
\label{s:Discussion}

We have presented a new gauge driver evolution system in
Sec.~\ref{s:GaugeDriverEquations} that makes it possible to impose a
wide range of gauge conditions in the generalized harmonic (GH) formulation
of the Einstein equations, without destroying its hyperbolicity.
The key idea is to construct an
auxiliary hyperbolic evolution equation for the gauge source function
$H_a$ that drives it toward the desired target $F_a$.  
Section~\ref{s:SpecificGauge} shows how many of the gauge conditions
widely used by the numerical relativity community can be included in
this way.  In Sec.~\ref{s:Stability} we analyze the effectiveness and
stability of the combined GH Einstein and gauge driver system for the
case of perturbations of flat spacetime.  This analysis shows that the
gauge driver equation effectively drives $H_a$ toward $F_a$, when $F_a$ 
is specified {\it a priori} as a function of the spacetime coordinates.  
We were somewhat surprised to find, however, that the
gauge driver system can be quite unstable when it is coupled to the GH
Einstein system.  We found that common gauge conditions like maximal
slicing and the $\Gamma$-freezing gauge conditions are unconditionally
unstable when implemented using our gauge driver equation.  This does
not imply of course that those conditions are unsuitable for use with
other forms of the Einstein system (like BSSN), just that they cannot
be implemented in a completely stable way in the GH Einstein system
coupled to the particular gauge driver equations introduced here.
Fortunately, we were able to find some of the commonly used gauge
conditions that can be implemented in this way: certain Bona-Mass\'o
slicing conditions and a commonly used form of the $\Gamma$-driver
shift conditions.  Our 3D numerical tests in
Sec.~\ref{s:NumericalTests} show that the gauge driver system can
impose these gauge conditions stably and effectively for the
evolutions of perturbed single black hole spacetimes.

There has been a great deal of discussion in the literature about the
formation of shocks when certain dynamical gauge conditions are
imposed~\cite{alcubierre97,alcubierre_masso98,Alcubierre2005}.  However, 
these discussions do not apply when those same gauge
conditions are imposed using a driver condition.  The gauge driver
system imposes the desired gauge condition only approximately,
not exactly.  At best, the desired gauge condition is imposed exactly only
asymptotically in time as the system approaches a
time independent equilibrium state,
and even in this state shocks do not necessarily form.  On the
contrary, there are many solutions even to bad gauge conditions that do
not have shocks.  What determines whether an evolution
system develops shocks is the structure of the operator that evolves
the spacetime metric and auxiliary fields.  Our evolution system (including
the gauge driver system) has been carefully designed to be linearly 
degenerate, a condition that prevents the formation of 
shocks (resulting from a crossing of characteristics)
from smooth initial data~\cite{liu1979}.  
Linear degeneracy does not prevent the formation of curvature
singularities, of course, or even the formation of coordinate
singularities that may arise from non-linearities in the non-principal
parts of the evolution equations.  

Causality is another issue that appears to be less restrictive for our
gauge driver system than it is for directly imposed gauge conditions.
For example, the parameter $\nu$ that appears in the $\Gamma$-driver
system discussed in Sec.~\ref{s:ShiftConditions} must take values
in the range $0 \leq \nu \leq \frac{3}{4}$ in order for that 
$\Gamma$-driver to evolve the
shift in a causal way in the BSSN system~\cite{Alcubierre2003a}.  
There is no such restriction on $\nu$, however,
when this $\Gamma$-driver is imposed through our gauge
driver system.  In our system the shift is evolved, along with the rest of
the spacetime metric, by the GH Einstein system.  This system is
manifestly hyperbolic, and all of the fields propagate within the
physical light cone, no matter what target gauge source function is
used in the gauge driver system.

It is easy to
imagine that the system presented here could be improved in several
ways.  It may be possible, for example, to improve the performance of
the system by formulating boundary conditions for $H_a$ that impose
the desired gauge condition $H_a=F_a$ exactly at the boundaries.  It
may also be possible to formulate a different evolution operator for
$H_a$ that drives it more stably and/or more efficiently toward the
desired target $F_a$.  Finally it may be possible to find better
target gauge conditions $F_a$.  The ones studied here are those which
have been found useful in evolutions of traditional three-plus-one
formulations of the Einstein system like BSSN.  But there may exist
gauge conditions having much better stability and effectiveness
properties when used as target gauge conditions within a gauge driver
system.  These questions, and others, will be addressed in future work
on this problem.


\acknowledgments We thank Harald Pfeiffer and Bela Szilagyi for helpful
comments concerning this work.  The numerical simulations presented
here were performed using the Spectral Einstein Code (SpEC) developed
at Caltech and Cornell primarily by Larry Kidder, Harald Pfeiffer,
and Mark Scheel.  This work was supported in part by grants from the
Sherman Fairchild Foundation and the Brinson Foundation, by NSF grants
DMS-0553302, PHY-0601459, PHY-0652995, and by NASA grant NNG05GG52G.

\bibstyle{prd}
\bibliography{../References/References}
\end{document}